\documentclass{aa}  
\usepackage{graphicx}
\usepackage{longtable}
\usepackage{txfonts}
\usepackage{natbib}
\usepackage{psfrag}
\def\bi {\begin{itemize}}
\def\ei {\end{itemize}}
\def\integral{\textit {INTEGRAL}}
\def\rxte{\textit {RXTE}}
\def\xmm{\textit {XMM}}

\def\sax{\textit {BeppoSAX}}
\def\swift{\textit {Swift}}
\def\thc {3C~273}

\def\fvar {$F_{\rm var}$}

\def\mm { millimeter }

\def\taumax {$\tau_{\rm max}$}

\begin{document}
   \title{The multiwavelength variability of 3C~273}

   \author{S. Soldi
          \inst{1,}\inst{2}
	 \and M. T\"{u}rler
	  \inst{1,}\inst{2}
	 \and S. Paltani
	  \inst{1,}\inst{2}	 
	 \and H. D. Aller
	  \inst{3}
	 \and M. F. Aller
	  \inst{3}
	 \and G. Burki
	  \inst{2} 
	 \and M. Chernyakova
	  \inst{4}
	 \and A. L\"ahteenm\"aki
	  \inst{5}
	 \and I. M. McHardy
	  \inst{6}
	 \and E. I. Robson
	  \inst{7}
	 \and R. Staubert
	  \inst{8}
	 \and M. Tornikoski
	  \inst{5}	 
	 \and R. Walter
	  \inst{1,}\inst{2}
	 \and T. J.-L. Courvoisier
	  \inst{1,}\inst{2}
          }

   \offprints{Simona.Soldi@obs.unige.ch}

   \institute{ISDC Data Centre for Astrophysics,
              Chemin d'\'Ecogia 16, 1290 Versoix, Switzerland
         \and Observatoire de Gen\`eve, University of Geneva, 
	      Chemin des Maillettes 51, 1290 Sauverny, Switzerland 
	 \and University of Michigan, Department of Astronomy, 
	     817 Dennison Building, Ann Arbor, MI 48\,109, USA
	 \and Dublin Institute for Advanced Studies, 
	      31 Fitzwilliam Place, Dublin 2, Ireland
	\and Mets\"ahovi Radio Observatory, Helsinki University of Technology, TKK,
	       Mets\"ahovintie 114, FIN-02540 Kylm\"al\"a, Finland
	\and School of Physics and Astronomy, University of Southampton, 
	       Southampton SO17 1BJ, UK
	\and UK Astronomy Technology Centre, Royal Observatory Edinburgh,
	       EH9 3HJ, UK
	\and IAAT, Abt. Astronomie, Universit\"at T\"ubingen, 
	       Sand 1, 72076 T\"ubingen, Germany
	      }

   \date{Received April 11, 2008; accepted May 16, 2008}

  \abstract
   {}
   {We present an update of the \thc's database hosted by the ISDC, completed with data from radio to gamma-ray observations over the last 10 years. 
   We use this large data set to study the multiwavelength properties of this quasar, especially focussing on its variability behaviour.}
   {We study the amplitude of the variations and the maximum variability time scales across the broad-band spectrum and correlate
   the light curves in different bands, specifically with the X-rays, to search for possible connections between the emission at different
   energies.}
   {\thc\ shows variability at all frequencies, with amplitudes and time scales strongly depending
   on the energy and being the signatures of the different emission mechanisms.    
   The variability properties of the X-ray band imply the presence of either two separate components (possibly a Seyfert-like and a blazar-like)
   or at least two parameters with distinct timing properties to account for the X-ray emission below and above $\sim$20 keV.
   The dominant hard X-ray emission is most probably not due to electrons accelerated by the shock waves in the jet as their variability does not 
   correlate with the flaring millimeter emission, but seems to be associated to long-timescale variations in the optical. 
   This optical component is consistent with being optically thin synchrotron radiation from the base of the jet and the 
   hard X-rays would be produced through inverse Compton processes (SSC and/or EC) by the same electron population.
   We show evidence that this synchrotron component extends from the optical to the near-infrared domain, where it is blended 
   by emission of heated dust that we find to be located within about 1 light-year from the ultraviolet source.}
   {}

   \keywords{Astronomical data bases: miscellaneous -- Galaxies: active -- quasars: individual: 3C~273 -- X-rays: galaxies -- Infrared: galaxies -- Ultraviolet: galaxies
               }

   \maketitle


\section{Introduction}
Active galactic nuclei (AGN) are known to emit a broad-band spectrum that extends
from radio to gamma-rays. Our understanding of their multiwavelength emission relies on the identification
of all components that contribute to the radiation.\\
\object{3C 273}, a bright and nearby ($z$ = 0.158) radio loud quasar, is a well known example of an AGN whose emission has been
extensively studied in the past at all frequencies \citep{courvoisier87,courvoisier90}. 
It shows most of the properties characteristic of blazars, like strong radio emission, a jet with apparent superluminal motion, 
large flux variations and, occasionally, polarisation of the optical emission \citep[for a review, see][]{courvoisier98}.
Synchrotron flares from a relativistic jet dominate the radio-to-millimeter energy output and extend 
up to the infrared and optical domains \citep{robson93,turler00}, whereas thermal emission 
from dust is at least in part responsible for the infrared quiescent continuum \citep{robson83,turler06}. 
The particularly bright excess in the optical-UV band has been interpreted as a signature 
of the accretion disc \citep[][and many subsequent studies]{shields78}, maybe in the presence of an external X-ray source 
\citep{collin91} or of a hot corona \citep{haardt94}, and seems to be due to the contribution
of two components with different variability properties \citep{paltani98a}. 
In the X-rays, it has been suggested that the soft-excess could be due to thermal Comptonisation of cool-disc photons 
in a warm corona \citep{page04}. 
Inverse Compton processes of a thermal plasma in the disc or in a corona and of a non-thermal plasma associated to the jet 
are believed to generate the X-ray to gamma-ray emission \citep{kataoka02,grandi04}. 

The \thc's Database hosted by the ISDC\footnote{\emph{http://isdc.unige.ch/3c273/} } is one of the most complete multiwavelength 
databases currently available for an AGN. 
When it was first published \citep{turler99a}, it contained 70 light curves covering 16 orders of magnitude in photon energy
and 30 years of observations.
We present here an update of the database with recent data and special emphasis on the X-ray emission.
Using these data, we analyse the multiwavelength variability properties of \thc, studying both the amplitudes and the time 
scales of the variations as a function of the frequency and trying to understand the different variability behaviours in the 
framework of already existing emission models.
We cross-correlate the fluxes in different energy ranges to investigate the connections between the different emission components, 
especially concentrating on the origin of the X-ray emission.
%
   \begin{figure*}[ht]
   \centering
   \includegraphics[angle=0,width=\textwidth]{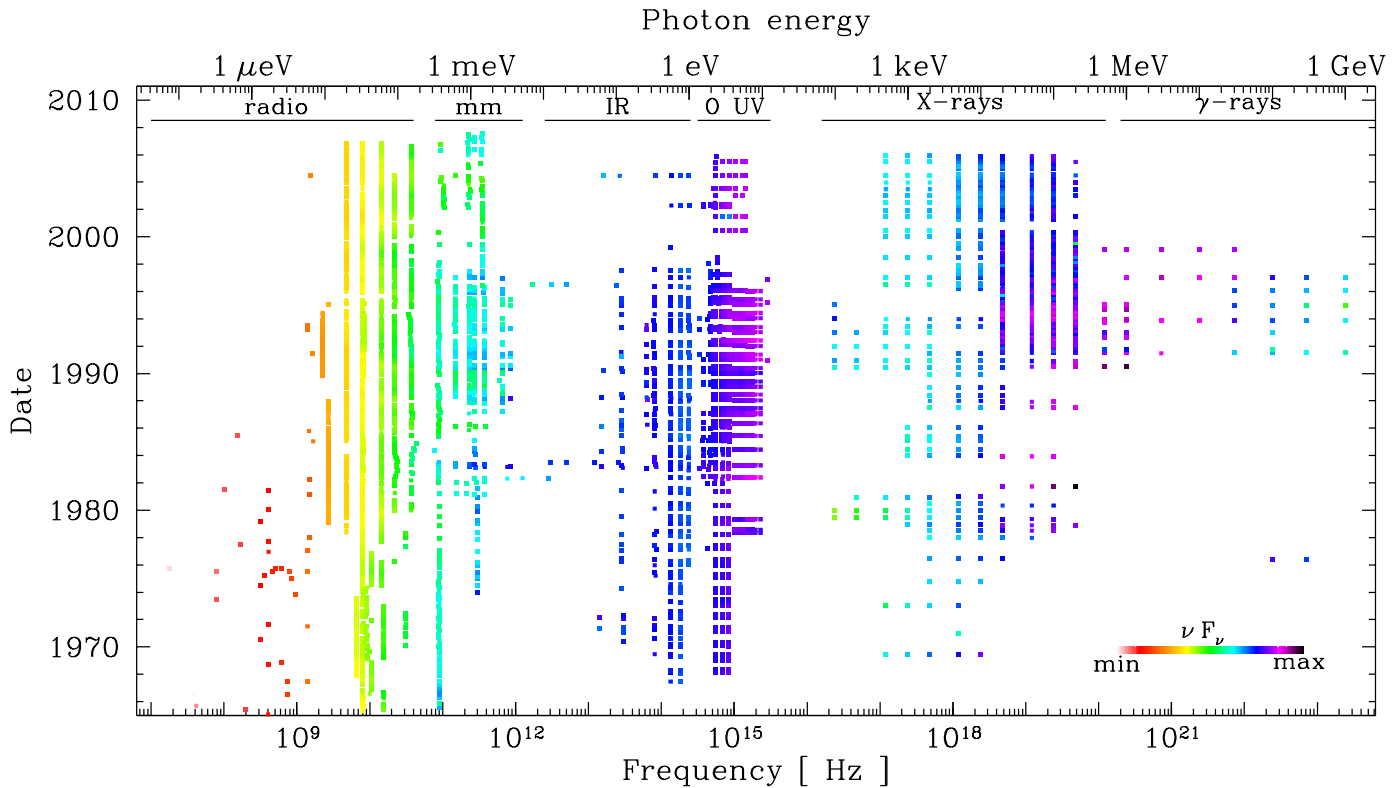}
      \caption{Time coverage of all observations of the database as a function of the frequency. The colour-scale
	 indicates the intensity of the observed flux in $\nu F_{\nu}$ representation and spans from a minimum intensity
	 of $6 \times 10^{-14}$ to a maximum of $1.6 \times 10^{-9}$ $\rm erg \, cm^{-2} \, s^{-1}$.
               }
         \label{fig:colorplt}
   \end{figure*}
\section{Data collection}
We included all the \thc\ public data of the last 10 years in the database (Table~\ref{table:var}), 
following the same criteria as \citet[][hereafter Paper I]{turler99a}.
Fig.~\ref{fig:colorplt} shows the time coverage of the data as a function of frequency.
%
\addtocounter{table}{1}
\subsection{Radio and millimeter data}
In the radio, we added the recently published data at 22 and 37\,GHz from the
Mets\"ahovi Radio Observatory \citep{terasranta04,terasranta05}, complemented by
unpublished data from this observatory and from the University of Michigan Radio
Astronomy Observatory (UMRAO) at 4.8\,GHz, 8.0\,GHz and 14.5\,GHz. We also added
the measurements from the 100\,m antenna at Effelsberg published by \citet{reich98}
and \citet{turler06}. Finally, we completed some historic
light curves with early data not published in tabular format by extracting them
from figures. This was done for the earliest Haystack data at 15.5\,GHz in
\citet{allen68}, the UMRAO data at 8.0\,GHz in \citet{aller67} and
some Crimean time-averaged data at 22\,GHz derived from \citet{efanov77}.

At millimeter and submillimeter (mm/sub-mm) wavelengths we added the observations
published by \citet{robson01} from the SCUBA instrument at 850\,$\mu$m on
the James Clerk Maxwell Telescope (JCMT), Hawaii and more recent unpublished
observations by this telescope. Other mostly unpublished observations are from the
Swedish-ESO Submillimeter Telescope (SEST) in La Silla, Chile, from the
Submillimeter Array\footnote{\emph{http://sma1.sma.hawaii.edu/tools.html}} 
(SMA) on Mauna Kea, Hawaii (for information on the quasar monitoring program at the SMA
see \citealt{gurwell07}), the Australia Telescope Compact
Array (ATCA), and the Owens Valley Radio Observatory (OVRO), California.

\subsection{IR data}
An important addition to the infrared dataset are the long-term light curves published by
\citet{neugebauer99} in the J, H, K, L and N bands from the Palomar Hale
5\,m telescope. Unfortunately, these data could not be obtained in tabular form
and have therefore been extracted from the figure and converted into flux
densities using the zero-magnitude fluxes given by \citet{neugebauer79}.
Other interesting additions are the data obtained during synchrotron outbursts
by \citet{mchardy99,mchardy07}, the data of April 2002 from \citet{jorstad04} 
and the data of June 2004 from \citet{turler06}. 
In the far-infrared, we added the observations by the Infrared Space
Observatory (ISO) published by \citet{haas03}.

The European Southern Observatory (ESO) magnitudes are now converted to flux using 
the zero-magnitude fluxes published by \citet{bouchet91} . 
To have the most coherent dataset with the observation of the United Kingdom Infrared 
Telescope (UKIRT), we now also rescaled the ESO data from \citet{litchfield94}
based on the ratio of ESO and UKIRT zero-magnitude fluxes.
Other data than those mentioned above have all been converted from magnitude to flux with
the UKIRT zero-magnitude fluxes that was used for all data in Paper I.

\subsection{Optical and UV data}
In the optical domain we included recent unpublished data in the Geneva
photometric system using the Mercator telescope on La Palma, Canary Islands.
We also added the observations of the Wise Observatory, Israel published by
\citet{giveon99} and \citet{kaspi00}. 
Based on 17 quasi-simultaneous (within 5 days) observations in the Geneva photometric system,
we derived a scaling factor of 1.106, applied to the B-band flux densities of \citet{kaspi00}.
In addition, we included the I and R-band data from \citet{jorstad04} 
and the R-band data from \citet{gosh00}.

The scaling of other data sets already included in Paper I did not change,
except for the Burkhead system \citep[][and references therein]{burkhead80} 
V-band magnitude correction that is now 0.00 instead of
$+$0.05. This difference comes from the correction of a quite obvious typo in
the publication of \citet{lyutyj76}: B-V should be +0.20 instead of +0.02 in line 8
of Table 1. In the R and I bands, we no longer use the zero-magnitude fluxes of
Paper I, but make a distinction, as far as possible, between data obtained in
the standard Johnson system (R: 2941\,Jy; I: 2635\,Jy) or the \citet{cousins76}
system (R: 3080\,Jy; I: 2550\,Jy). In particular we used the Cousins values for
the R-band magnitudes from \citet{giveon99}.

We also included V-band measurements by the Optical Monitoring Camera (OMC)
aboard the INTEGRAL satellite taken from the OMC
Archive\footnote{\emph{http://sdc.laeff.inta.es/omc/}}. We averaged contiguous
observations and converted magnitudes to fluxes in the Geneva photometric system as 
described in Paper I, without any additional correction ($V_{\rm corr} = 0 \,$, see Paper I).

Optical and UV data from \xmm/OM observations have been taken from \citet{chernyakova07}, 
when available, or extracted following the procedure reported there.
The 45 OM fluxes, covering from 7 to 12 different epochs, depending on the wavelength,
have been included in the database after checking that there are no calibration problems
and that the wavelengths of the 6 OM filters are sufficiently close to those of the light curves in the database.

Among the few observations performed with the \swift/UVOT monitor, only data collected with the V 
filter are included in our database, because, due to the brightness of \thc\  (m = 11--13 depending on the filter), the detector is operating 
above its saturation limit with all filters except V.
The V fluxes have been obtained analysing the data with the XRTDAS software v.2.5a distributed with the HEAsoft 6.1.1 
package and using the latest calibration files (2007-03-30) available at the time of the work.

Finally we included in the database an additional light curve in the far-UV at a wavelength of 1050 \AA\ 
with data extracted from \citet{appenzeller98} and \citet{kriss99}.
\subsection{X-ray data}
During the last 10 years, several X-ray satellites have extensively observed \thc, considerably increasing the amount of 
data available above 0.5 keV (Fig.~\ref{fig:colorplt}). In order to obtain a data set as uniform as possible, 
we fitted the available spectra in similar energy ranges using a power law model absorbed by a Galactic column density 
$N_{\rm H} = 1.8 \times 10^{20} \, \rm cm^{-2}$.
Some simultaneous or quasi-simultaneous observations (within about one day) of \thc\ have been performed in the X-rays
allowing the inter-calibration of the different instruments. 
{\it Chandra} observations of \thc\ were mainly concentrated on the X-ray jet \citep{sambruna01, jester06} 
and therefore have not been included in the database.
\begin{table*}
\caption{Inter-calibration constants for the X-ray light curves} 
\label{table:renorm} 
\begin{center}
\renewcommand{\footnoterule}{}
\begin{tabular}{c c c c c c c c}       
\hline\hline                
Energy [keV] & \rxte\ & \sax/LECS & \sax/MECS & \sax/PDS & \xmm/PN & \integral\ & \swift/XRT \\   
\hline                      
   0.5 & -               & 1.67 $\pm$ 0.07 & -               & -  & 1  & -  &  4.87$^a$ \\
   1   & -               & 1.35 $\pm$ 0.02 & -               & -  & 1  & -  &  4.87$^a$ \\
   2   & -               & -               & 0.87 $\pm$ 0.02 & -  & 1  & -  &  4.87$^a$ \\
   5   & 0.77 $\pm$ 0.12 & -               & 1               &  - & 1  & 1 &  4.87$^b$ $\pm$ 1.27 \\
   10  & 0.78 $\pm$ 0.15 & -               & 1               &  - & 1  & 1 &  - \\
   20  & 0.78$^c$  & -  & -  & 1  & -  & 1 &  - \\
   50  & 0.78$^c$        & -               & -               & 1  & -  & 1 &  - \\
   100 & 0.78$^c$        & -               & -               & 1  & -  & 1  &  - \\
   200 & -               & -               & -               & 1  & -  & 1  &  - \\
\hline                         
Renormalised to & \xmm/PN & \xmm/PN & \xmm/PN & \rxte & - & \rxte & \rxte \\   
\hline                         
\end{tabular}
\end{center}

$^a$ Fixed to the value found at 5 keV.\\
$^b$ This large scaling factor is due to the pile-up affecting the XRT images of \thc\ and it does not have to be considered as an inter-calibration constant between \swift/XRT and \rxte. \\
$^c$ Fixed to the value found at 10 keV.
\end{table*}
\subsubsection{{\it XMM-Newton}/PN data}
Among the numerous observations of \thc\ performed by \xmm, 16 sets of EPIC/PN data collected from June 14, 2000 to
July 10, 2005 were publicly available at the time of this work. The EPIC spectra
have been taken from \citet{stuhlinger04} and from \citet{chernyakova07}.
Due to their higher signal-to-noise ratio, we choose PN spectra instead of MOS1/2 for the spectral analysis.
The spectra are fitted separately in the ranges 0.5--2 and 2--10 keV in order to obtain fluxes and spectral indices
in the same energy bands as for \sax\ data.
\subsubsection{{\it RXTE} data}
Several campaigns have been performed with \rxte\ to monitor the X-ray emission of \thc.
We have collected all the data available at the time of this work in the \rxte\ archive, which consist in
957 observations performed between February 9, 1996 and December 11, 2005.
We use the \rxte\ standard products to derive a flux and a spectral index associated with each observation.
The HEXTE data show a low signal to noise ratio and consequently a large scatter in the resulting parameters when
fitted alone. Therefore we perform a PCA-HEXTE combined fit for each observation, resulting in a broad-band 
spectrum from 3 to 100 keV. Our choice is supported by the absence of strong features, like Compton reflection,
in the X-ray spectrum of \thc\ \citep{grandi04}. 

A slow gain drift of about 1\% over 2 years\footnote{\emph{http://heasarc.nasa.gov/docs/xte/e-c\_table.html}}
determines small changes in the energy-channel relation of the PCA instrument. Consequently, the maximum energy of channel 2 
moved from 2.02 to 2.96 keV over the mission time. Since the inclusion of this channel results in very high reduced chi-square
values $\chi^2_{\rm red}$, we exclude the data below 3 keV from all PCA spectral fitting.

We choose \xmm/PN as the reference instrument and renormalise the \rxte\ fluxes to the PN ones at 5 and 10 keV, obtaining
an inter-calibration factor of $\sim 0.78$. The light curves of the other X-ray instruments are then rescaled either to the \xmm/PN one 
or to the \rxte\, one, depending on the time frame and on the energy band. 
The inter-calibration factors reported in Table~\ref{table:renorm} for \sax/PDS, \integral\ and \swift/XRT are relative to
\rxte, therefore they are multiplied by $0.78$ to derive the final scaling factors (relative to \xmm/PN) 
that we apply for the light curves in the database.
\subsubsection{{\it Beppo-SAX} data}
The LECS, MECS and PDS spectra of the 9 \sax\ 
observations of \thc\ have been retrieved from the ASDC archive\footnote{\emph{http://www.asdc.asi.it/}} and separately fitted in the ranges
0.5--2 keV, 2--10 keV and 15--200 keV, respectively. This allows us to have a separate representation of the soft-excess region, 
the medium and the hard X-ray bands.
The 0.5--10 keV fluxes are renormalised to the \xmm/PN ones. The inter-calibration factor between the 20--200 keV fluxes from PDS
and the \rxte\ ones is found to be consistent with 1.
\subsubsection{{\it INTEGRAL} data}
From the beginning of the mission, \integral\ spent about 2 Ms on \thc, more than 1 Ms of which
(January 2003--July 2005) are publicly available and included in the database. Data analysis is described in detail in 
\citet{chernyakova07}.
Different observations were summed up in order to increase the statistics of the data and resulted in 5 sets of IBIS/ISGRI
and JEM-X data and 3 of SPI data. For each data set a combined fit with all the available high-energy \integral\ instruments 
is performed in the range 5--200 keV. 
SPI (or ISGRI when SPI data are not available) is assumed to be the most stable \integral\ instrument and the other 
inter-calibration factors are fixed to the values reported by \citet{chernyakova07}.
The inter-calibration factor between \integral\ and \rxte\ fluxes is consistent with 1.
\subsubsection{{\it Swift}/XRT data}\label{swift_data}
\swift\ has observed \thc\ several times, collecting data with all 3 instruments on board.
We have analysed and included in our database the 5 XRT observations performed in Photon Counting readout mode with
exposure time greater than 3 ksec (between November 15 and December 19, 2005).
The data were analysed with the XRTDAS software v.2.5a and the 2007-03-30
calibration files. During all these observations the source shows a count rate higher than 1 count/s and 
the image is affected by photon pile-up. In order to correct for this effect, the spectra are extracted in an annular region
of 6 and 20 pixels inner and outer radii, respectively \citep{vaughan06}. Therefore, only a fraction of the flux
can be recovered, which is estimated thanks to quasi-simultaneous \rxte\ observations (Table~\ref{table:renorm}).
For December 19, 2005 observation, we performed an exposure map correction on the data (using {\it xrtexpomap}) in order 
to account for the decreased collecting area due to dead columns of pixels crossing the source image. The effect is found to be 
negligible, showing no changes in the fitting parameters compared to the uncorrected data.
The fitting on all the XRT spectra is separately performed in the 0.5--2 and 2--8 keV energy bands.
\subsubsection{BATSE data}
In the hard X-ray range, the BATSE data from the Compton Gamma-Ray Observatory
(CGRO) have recently been made available by the BATSE Earth occultation
team\footnote{\emph{http://f64.nsstc.nasa.gov/batse/occultation/}} for four different
energy bands: 20--40, 40--70, 70--160 and 160--430\,keV. 
In order to make them fit the database format, we extracted
four different BATSE light curves for energies of 20, 50, 100 and 200\,keV
assuming a fixed power law photon index of $\alpha=1.7$. No rescaling factor has been applied to the BATSE light curves.
The variability analysis was performed on the 20--70 and 70--430 keV light curves separately
from the other hard X-ray data (see Sect.~\ref{sect_var}).
\subsection{Gamma-ray data}\label{gammaray_data}
In the gamma-ray domain, we included the combined 3\,MeV-to-10\,GeV power-law fits to the 
{\textit CGRO}/COMPTEL and EGRET data published by \citet{collmar00b} from 1996--1997 observations,
as well as single-energy-bin COMPTEL data from \citet{collmar00a,collmar00b}.
Additional {\textit CGRO}/EGRET data come from spectral fits published by \citet{mukherjee97}
and \citet{nandikotkur07}.
Finally, at 100 MeV, we included the EGRET fluxes at energies $E>100$ MeV given by \citet{mukherjee97} 
from observations in 1991--1994.
We note that there are sometimes several entries in the database for the same observations
analysed by different authors that can either add or reduce intrinsic source variability.
%
\section{Database format}\label{new_format}
We took the opportunity of the update of 3C 273's database to reorganise
slightly the data in different light curve files and to include additional
information. 
The database, available at \emph{http://isdc.unige.ch/3c273/}, is organised in 70 ASCII files, 
each of them containing a light curve at a certain wavelength. 
Each file is composed by 10 columns with one row per observation.
Most of these columns (date in decimal year, frequency, wavelength, flux, flux error,
reference) have maintained the same format as in Paper I.
The date in decimal year is still linearly related to the Julian Date,
but we now also give the date in UTC format as \texttt{yyyy-mm-ddThh:mm} instead
of the previous format \texttt{yyyymmdd.hh}.
For observations lasting several days, we express the date as a range of the form 
\texttt{yyyy-mm-dd/yyyy-mm-dd} in UTC format and with the decimal year corresponding 
to the middle of the observation.
In some cases when the day of the month or even the month is not known we cut
the UTC format to the last relevant number and assume the middle of the month
(day 15) or of the year (July 1st) to derive the decimal year.

To allow an easy selection of the data, we introduced a ``Flag'' column giving
information on the quality and reliability of the data. It can have integer
values from $-$3 to $+$1. The default value for good data points is zero and
less useful or reliable data points have negative flag values. A flag of $-1$
indicates ``useless'' data in the sense that they are compatible with
contemporaneous observations with smaller uncertainties, but their inclusion
just adds noise to the light curves. A flag of $-2$ indicates ``uncertain''
data that are less reliable because of large uncertainties or due to an
automatic processing without manual checking. Finally, a flag of $-3$ is
assigned to ``dubious'' data points that are clearly outliers and
incompatible with contemporaneous observations or with the overall variability
trend. Finally, we assign a flag of $+$1 to good infrared and optical 
data with a significant contribution by synchrotron flares lasting several days to weeks.

The flagging of the data was usually done manually by careful inspection of the
light curves. An automatic procedure was only used for the daily radio 
measurements from the Green Bank Interferometer (GBI). This was done for points
that were outstanding by more than 3-$\sigma$ from the 30-day smoothed
light curve as we do not expect true variability on a much faster time-scale than
a month at these low radio frequencies. For these data, as well as for the Earth
occultation measurements by {\textit CGRO}/BATSE in the hard X-rays, we both include the
single, daily observations with a flag set to $-$1 and the more easily usable
30-day averages with a flag of zero.

The meaning of the Flag value and additional information is included with each
measurement in a new column called ``Comments''. For instance, we included there
the scaling factor used for the GBI data according to the values given by
\citet{turler00}, as well as information on the assumed hydrogen column
density $N_{\mathrm{H}}$ (fixed or free) for X-ray data from the literature.
Finally, the information on the telescope or the antenna in the ``Observatory''
column was also extended in particular by the addition of its diameter in meter.

In addition, at the same web page, for each light curve a figure in GIF format is now provided,
where the data points with different flag values are indicated by different colours.
%
\section{Variability analysis}\label{sect_var}
In order to use only the best-quality data for the variability analysis, we consider only Flag $\geq 0$ data and
exclude the light curves with less than 10 data points. 
Moreover, due to inter-calibration problems, the radio light curves at $\nu \le 1.5$ GHz and the I and R optical data sets are not
included.

\begin{figure}[!t]
   \centering
   \includegraphics[angle=0,width=9.0cm]{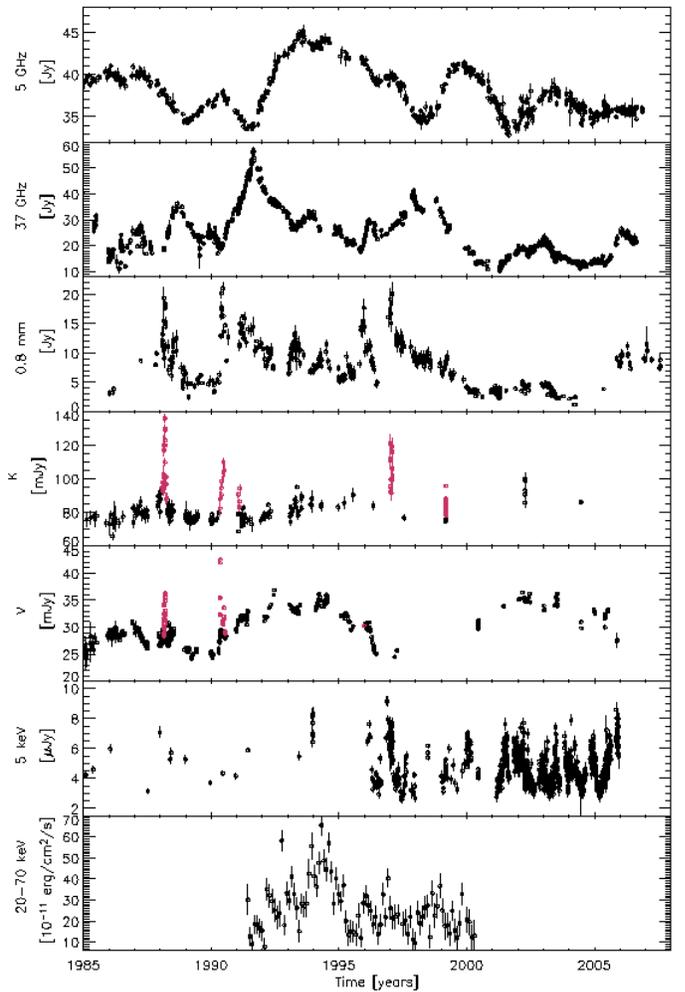}
      \caption{Examples of light curves from the 3C~273 database for the last 23 years of observations,
      at 5 GHz, 37 GHz, 0.8 mm, in the K band, in the V band, at 5 keV and in the 20--70 keV range (this latter rebinned to 1-month bins).
      The data during strong synchrotron flares (Flag = 1) are indicated in grey (red in the electronic version) for the optical and IR data sets
      [for a higher resolution figure, see \emph{http://isdc.unige.ch/$\sim$soldi/image/9947F2\_hr.ps}].
               }
         \label{fig:lc}
   \end{figure}

Synchrotron flaring emission is known to contribute in the infrared-optical domain \citep{courvoisier88}. 
In order to separately study the variability behaviour of the underlying components, infrared and optical light curves 
excluding the periods of known synchrotron flares (January--June 1972, December 1982 -- June 1983, January--May 1988,
February--August 1990 and January--March 1997, flagged with 1 in the database; see Fig.~\ref{fig:lc}) 
are also analysed (`wf' data sets in Table~\ref{table:var}).

The 20--70 and 70--430 keV BATSE light curves are used separately from the other X-ray data, as they have been rebinned 
to about one-month bins in order to have an acceptable signal-to-noise ratio for the variability analysis (Fig.~\ref{fig:lc}).

The \rxte/PCA fluxes included in the database have been extracted from the spectra accumulated during observations
from 0.2 to 17 ks long. In order to study variability on scales as short as possible  we use the PCA light curves (in counts/s) provided 
by the \rxte\ archive in the 4--9 and 9--20 keV ranges, after rebinning them to 1000-second bins. 

\begin{figure}[!b]
   \centering
   \includegraphics[angle=90,width=9.3cm]{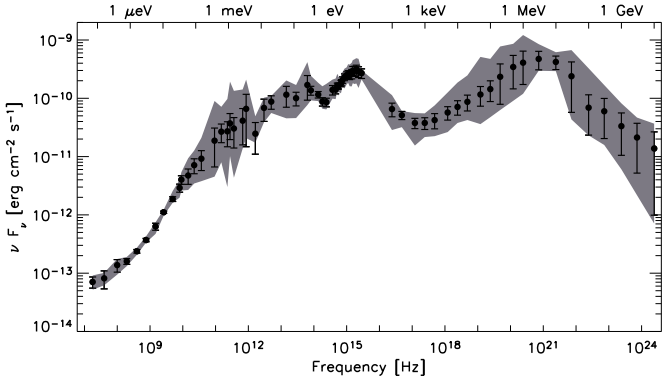}
      \caption{Average spectral energy distribution of \thc\ from radio to gamma-rays (points), spanning from 4 to 44 years of observations
      depending on the wavelength. The error bars are calculated as the standard deviation from the mean values and the grey area 
      represents the observed range of variations. 
               }
         \label{fig:sed}
   \end{figure}

In Table~\ref{table:var} the light curves are reported together with the time range between the first and last observation included 
in the database, the number of data points, the average frequency, the average
flux with its standard deviation, the $\nu \rm F_{\nu}$ intensity and the variability parameters that will be discussed in the following sections.
Some examples of light curves from the database are shown in Fig.~\ref{fig:lc}.
The average spectral energy distribution (SED) including all the light curves of the database is shown in Fig.~\ref{fig:sed}.
Beside the synchrotron and Compton peaks, typically observed in the SED of blazars, 
\thc\ shows a prominent excess in the optical-UV band, the so-called blue bump.
\subsection{Fractional-variability amplitude}
   \begin{figure}
   \centering
   \hspace{-0.8cm}
   \includegraphics[angle=90,width=9.3cm]{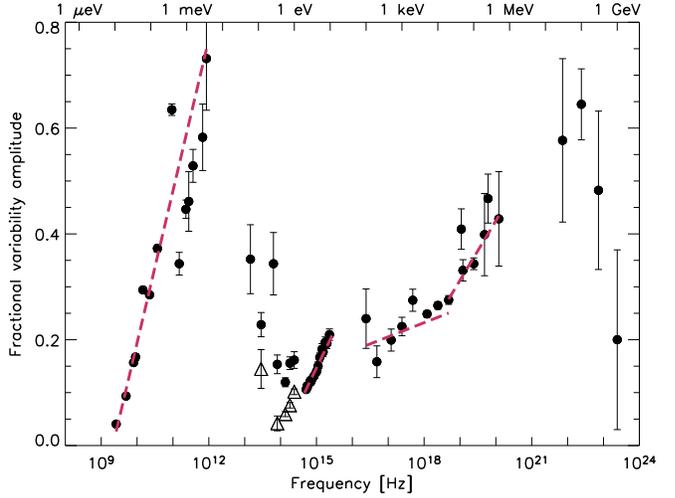}
      \caption{Fractional-variability, \fvar, spectrum of \thc\ from radio to gamma-rays (filled circles). 
       Open triangles have been obtained after removal of flares in the infrared band.
       The dashed lines in the radio-millimeter, optical-UV and X-ray bands represent the best fits.
               }
         \label{fig:Fvar}
   \end{figure}
\subsubsection{Definition}\label{def_fvar}
AGN are known to be very variable objects at all wavelengths and different methods can be used to
quantify the amplitude of the variations.
In order to estimate the total variability of each light curve in our database we use the fractional
variability amplitude, $F_{\rm var}$, given by \citep[e.g.][]{edelson02}:
\begin{equation}
F_{\rm var} = \sqrt{\frac{S^2-\overline{\varepsilon^2_{\rm err}}}{\bar{x}^2}}  ,
\end{equation}
where $S^2$ is the sample variance of the light curve, $\bar{x}$ is the average flux and 
$\overline{\varepsilon^2_{\rm err}} = \frac{1}{N}\sum_{i} \varepsilon_{\rm err,\it i}^2$ is the mean of the squared measurement uncertainties.
\fvar\ is calculated for all the light curves in our sample (Table~\ref{table:var}) and the resulting spectrum, from radio to gamma-rays, 
is shown in Fig.~\ref{fig:Fvar}.
The \fvar\ values obtained when excluding synchrotron flares from IR and optical light curves are also calculated,
though in the optical range the flares are too few to have any significant influence on the value of \fvar .

The uncertainty on \fvar\ due to the measurement-error fluctuations have been estimated through Monte Carlo simulations 
by \citet{vaughan03} (see Eq.~(B2) there).
Applied to our data sets, it results in percentage uncertainties smaller than 5\% in more than 80\% of the light curves.
In order to give a more conservative estimate of the uncertainty we use a bootstrap procedure
resulting in error bars that are in average twice as large as those proposed by \citet{vaughan03}.
For each frequency, this procedure consists in randomly drawing a number of points $N$ from the corresponding light curve,
without avoiding repetitions and $N$ being the total number of data points in the original light curve.
Then the \fvar\ parameter is calculated on the simulated light curve. This is repeated a statistically significant number of times 
(100,000 here), in order to obtain a distribution of the \fvar\ values found. A Gaussian fit is then applied and the standard 
deviation obtained is taken to represent the uncertainty of \fvar\ for the corresponding original data set.
%
\subsubsection{Results on the variability amplitude}\label{results_fvar}
The fractional variability amplitude strongly depends on frequency, as shown in Fig.~\ref{fig:Fvar}.
Even though different in slopes, the radio-mm, optical-UV and X-ray \fvar\ all show a clear increasing trend.
In particular the increase of \fvar\ is linear with the logarithm of the frequency 
with a slope of $0.287 \pm 0.004$ and $0.14 \pm 0.01$ in the radio-mm and optical-UV bands, respectively (Fig.~\ref{fig:Fvar}). 
In the X-rays, a single linear increase from 0.1 to 500 keV does not represent the data well.
There seem to be rather two different behaviours below and above $\sim$ 20 keV, with a steeper slope of
$0.11 \pm 0.02$ at higher energies when compared to the much flatter trend in the lower band ($0.03 \pm 0.01$).

The steep increase of \fvar\ from radio to millimeter is well understood in the frame of the evolution of 
synchrotron flares that are believed to propagate from higher to lower energies, with increasing
duration \citep{turler00}. As a consequence, consecutive flares overlap, 
creating an intense, continuous radio emission with relatively small variations, 
whereas in the \mm\ band very short flares stand out on a lower quiescent emission, resulting 
in a large fractional variability amplitude (Fig.~\ref{fig:lc}).

The position of the \fvar\ peak in the mm band, around 0.35 mm ($8.6 \times 10^{11}$ Hz), is uncertain due to the scarcity 
of data in the far-IR range. Nevertheless, a clear decreasing trend of \fvar\ is visible in the IR band.  
In the 1--10 $\mu$m light curves (J to N) where the synchrotron flares have been subtracted, 
\fvar\ decreases by a factor of 2 or more compared to the same data sets including the flares.
In particular, the variance of the M curve is smaller than the squared measurement 
uncertainties, whereas not enough points are left in the Q light curve to estimate \fvar.
In the near-IR around 4 $\mu$m (L band), \fvar\ again starts to increase towards shorter
wavelengths, still staying below 0.1.
The less variable component arising in the mid-IR band is likely to be thermal emission from dust. 
As it approaches the sublimation temperature, the dust emission is contributing decreasingly, therefore
another component has to be responsible for most of the observed radiation and variability at wavelengths around 1 $\mu$m (J and H bands).

The fractional variability amplitude in the optical--UV domain steadily increases from 0.1 to 0.2.
This trend is consistent with the decomposition suggested by \citet{paltani98a}, in which
the red (`$\mathcal{R}$') component, dominant in the optical range, shows smaller variations than the 
blue (`$\mathcal{B}$') component, which is dominant at UV frequencies, the total emission resulting from changes in the 
relative fluxes of these two components. 

   \begin{figure}
   \centering
   \includegraphics[angle=90,width=8cm]{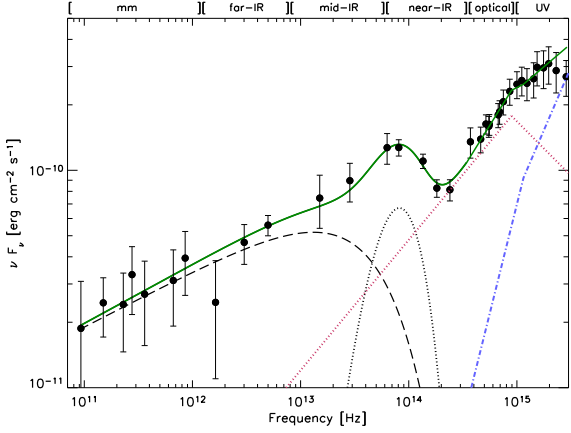}
      \caption{Average mm to UV spectral energy distribution. The best fit model to the mm-IR data (continuous line) is obtained 
      adding a cut-off power law (dashed line), a grey-body dust contribution (dotted curve) and the broken-power-law $\mathcal{R}$ 
      component from \citet{paltani98a} (dotted lines).
      In the optical-UV band the continuous line is the sum of the $\mathcal{R}$ and $\mathcal{B}$ (dot-dashed line) components.
      The mm to optical fluxes refer to the light curves where synchrotron flares have been subtracted. 
                     }
         \label{fig:sed_IR}
   \end{figure}
Considering the continuous trend of \fvar\ from $8 \times 10^{13}$ to $3 \times 10^{15}$ Hz, $\mathcal{R}$ could extend towards long
wavelengths to the near-infrared domain and be responsible for most of the emission and variability properties in this range, outside 
the short-flare periods.
The synchrotron emission, as extrapolated from the mm band, and the dust explain the far- and mid-IR emission, 
but a cut-off or a break is required in the synchrotron radiation to avoid overestimating the near-IR flux.
We thus fit the mm to IR spectrum (excluding synchrotron flares) with an exponentially cut-off power law 
($k_{\rm mm} \, \nu^{-\alpha_{\rm mm}} e^{-\frac{\nu}{\nu_{\rm c}}}$) and fix the $\mathcal{R}$ contribution as modelled by \citet{paltani98a}.
Paltani et al. found that the optical-UV data were consistent with $\mathcal{R}$ having either a single ($\equiv \mathcal{R}_{\rm max}$ 
case, with a slope of --0.7) or a broken power law shape ($\equiv \mathcal{R}_{\rm min}$ case, with slopes of 0.0 and --2.9). 
We choose to model $\mathcal{R}$
here with a broken power law with slopes of --0.4 and --1.5, before and after the break, as this represents an intermediate
shape between the $\mathcal{R}_{\rm min}$ and $\mathcal{R}_{\rm max}$ extreme cases.
The fit results in a reduced chi-square $\chi^2_{\rm red} = 1.5$, mainly due to the deviation of the near-IR points from the model.
Adding a component to account for the dust emission improves significantly the fit ($\chi^2_{\rm red} = 0.36$, 
F-test probability = $2 \times 10^{-4}$) and gives $\alpha_{\rm mm} = 0.74 \pm 0.06$, $k_{\rm mm} = 1.9 \times 10^{-11}$ at $10^{11}$ Hz 
and $\nu_{\rm c} \simeq 5 \times 10^{13}$ Hz (Fig.~\ref{fig:sed_IR}).
The dust contribution is represented by an isothermal grey-body\footnote{An isothermal grey-body at temperature $T$ emits at the observed frequency 
$\nu_{\rm obs}$ a flux density given by:
\begin{equation}
F_{\nu}^{\rm obs} (\nu_{\rm obs}) = (1+z) \, A_{\rm dust} \, D_{\rm L}^{-2} (1-e^{-\tau_{\nu}}) \, B_{\nu}^{\rm em} (\nu_{\rm em}, T)
\end{equation}
where $B_{\nu}^{\rm em} (\nu_{\rm em}, T)$ is the Planck function for a black body at temperature T, $D_L$ is the luminosity distance,
$A_{\rm dust}$ is the area of the projected source, $\tau_{\nu}$ is the optical depth that can be expressed as a function of the dust emission
index $\beta$ and the frequency at which the optical depth equals 1, $\tau_{\nu} = (\nu_{\rm em}/\nu_{\tau =1})^{\beta}$.
This model does not take into account the source geometry \citep{polletta00,turler06}.
}
with fitted temperature $T_{\rm dust} = 1156 \pm 92$K and emitting area
$A_{\rm dust} = 1.5 \pm 0.5 \rm \, pc^2$. The other grey-body parameters (i.e. the dust emissivity index and the frequency at
which the optical depth equals 1) are chosen as in \citet{turler06}. 
We apply the same model to the spectrum that includes synchrotron flares and to that derived using only synchrotron flares 
and we find that the same model with very similar parameters can represent the different states of the mm to IR emission.
The small differences between the $T_{\rm dust}$ and $A_{\rm dust}$ values we find and those reported by \citet{turler06} 
could be due to variability of the dust component, but can also be explained by the strong dependence of the dust modelling on the extrapolation 
of the mm emission which is quite uncertain due to the large variability of this energy band. 

In the X-ray domain, there is an increase of the fractional variability amplitude from 0.2 to 500 keV.
However, the 0.1--0.5 keV curves are extremely sensitive to the addition or subtraction
of a few data points, making highly uncertain the \fvar\ trend in this band.
We can exclude the variations below 10 keV being due to the hydrogen column density, $N_{\rm H}$,
because this would imply extremely larger variations at low energies than are observed. 

For some gamma-ray light curves \fvar\ cannot be calculated because of the small amount of available data 
or the large measurement uncertainties (1--10 MeV, 3--10 GeV). 
The increase of \fvar\ observed in the X-rays continues up to the gamma-rays, possibly even above 100 MeV, because
the observed decrease in the GeV range is most likely due to very large flux uncertainties.
In addition, caution should be used when considering the variability results in the gamma-ray band above 30 MeV, 
because these light curves include also fluxes obtained from the same observations by different analyses 
(see Sect.~\ref{gammaray_data}).
%
\subsection{Characteristic time scales of variability}\label{timescale}
\subsubsection{Method}\label{timescale_method}
\begin{figure}
   \centering
   \hspace{-0.8cm}
   \includegraphics[angle=90,width=7cm]{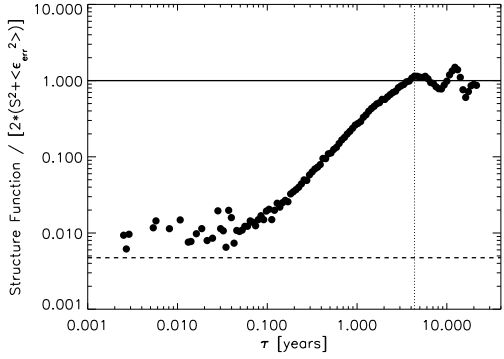}
      \caption{Example of structure function for the 8 GHz light curve, renormalised to the expected value of the upper plateau,
		($2 \cdot (\rm S^2 + \overline{\varepsilon_{\rm err}} ^2)$, continuous line). The dashed line is the expected value of 
		the lower plateau ($2 \cdot \overline{\varepsilon_{\rm err}} ^2$). The maximum time scale of variability is found 
		to be around 4.4 years (vertical dotted line), when the SF flattens reaching the upper plateau.
               }
         \label{fig:sf_8GHz}
   \end{figure}

A method to estimate the characteristic time scale of a light curve is given by the structure function (SF).
For a time series $F(t)$, the SF is defined as a function of the time lag $\tau$ by \citep{simonetti85}:
\begin{equation}
\rm SF(\tau) = \it <(F(t+\tau) - F(t))^2> \,\,\,\, .    \label{equ_SF}
\end{equation}
The SF is initially flat at a level given by the average measurement uncertainty, $2 \cdot \overline{\varepsilon_{\rm err}} ^2$, it then
increases up to a maximum variability time scale, $\tau_{\rm max}$ , indicating that the source variations are correlated.
For $\tau > \tau_{\rm max}$, the SF flattens again at a value corresponding to $2 \cdot (\rm S^2 + \overline{\varepsilon_{\rm err}} ^2)$,
where $S^2$ is the variance of the data (see an example in Fig.~\ref{fig:sf_8GHz}).

We apply the structure function analysis to all data sets with at least 200 points and then further select those for which
the upper plateau is clearly present, allowing a measurement of $\tau_{\rm max}$ (Fig.~\ref{fig:taumax}).
In a few cases, for the 0.8 mm and some of the optical light curves, the SF keeps increasing 
beyond the $2 \cdot (\rm S^2 + \overline{\varepsilon_{\rm err}} ^2)$ value, indicating a $\tau_{\rm max}$ longer than 
the sampled time scales (i.e. $\gtrsim$ 9 years).
We exclude the light curves that have SFs for which it is difficult to reliably estimate the position of
the plateau and the corresponding $\tau_{\rm max}$.

In order to estimate the uncertainty on \taumax, for each data set we simulate 1,000 light curves (with a bootstrap procedure), 
compute the SFs and calculate \taumax. To automatize the process, we define \taumax\ as the shortest time scale for which the SF is 
higher than 97\% of the expected value of the upper plateau. 
Then the distribution of the simulated \taumax\ values is calculated and
the uncertainty on the original \taumax\ is calculated as the, possibly asymmetric, 1$\sigma$ standard deviation from the median value.
%
\subsubsection{Results on the variability time scales}
   \begin{figure}
   \centering
   \hspace{-0.8cm}
   \includegraphics[angle=90,width=8.7cm]{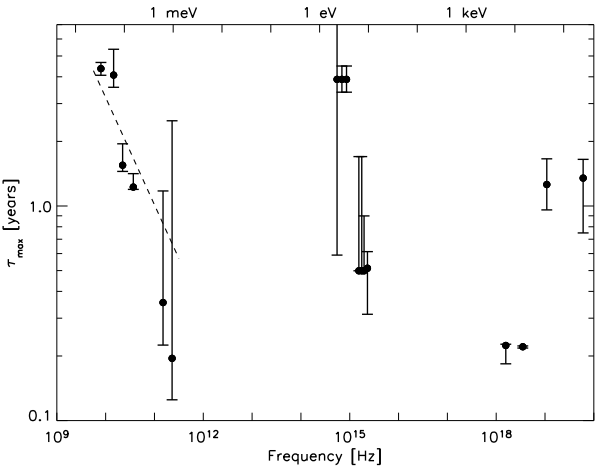}
      \caption{Spectrum of the maximum variability time scale \taumax\ estimated from the structure function analysis 
      	       (the lower limits are not displayed).
      	       The error bar calculation is explained in Sect.~\ref{timescale_method}.
	       The dashed line represents the best fit to the radio-mm data using Eq.~(\ref{t_cool}).
               }
         \label{fig:taumax}
   \end{figure}
The maximum time scales of variability from radio to mm frequencies are interpreted as due to the typical 
evolution of synchrotron flares, showing longer duration.
The 22 and 37 GHz values of \taumax, calculated with a structure function analysis, have been already reported by \citet{hovatta07}.
Assuming that the measured variability time scales in the radio-mm band represent an upper limit to the cooling time of electrons through
synchrotron emission:
\begin{equation}
\tau_{\rm max}(\nu) \ge t_{\rm cool} \simeq  6 \times 10^8 \, B^{-3/2} \, \nu^{-1/2} \, D^{-1/2} ,   \,\,\,\,  \rm s \label{t_cool}
\end{equation}
 (where $t_{\rm cool}$ is in seconds, $B$ in Gauss, $\nu$ in MHz and $D$ is the jet Doppler factor), we can obtain a lower limit to the magnetic field responsible 
for this emission. This is just a rough lower limit of $B$ as Eq.~(\ref{t_cool}) is obtained in the case of a plasma cloud cooling through synchrotron 
mechanism in a constant magnetic field, a rather different condition than that of a shock in a jet.
The fit of $\rm Log (\tau_{\rm max})$ vs $\rm Log (\nu)$ with Eq.~(\ref{t_cool}) results in $B \ge B_{\rm min} = 0.089 \pm 0.003$ Gauss
 (Fig.~\ref{fig:taumax}), when $D = 5$ \citep[][and references therein]{savolainen06}.
This result is consistent with the magnetic field range found by previous studies, where an estimate 
of $B \sim 0.4-1$ Gauss was obtained using the time scales of sub-mm to cm and IR flares \citep{courvoisier88, robson93}. 

In the UV range, \taumax\ is basically constant (as already found by \citealt{paltani98a, favre05}) 
and much shorter than in the optical range, where \taumax\ is about 4 years or even longer
than the sampled time scales.
This is due to the dominance in the UV of the $\mathcal{B}$ component \citep{paltani98a}.
   \begin{figure}
   \centering
   \hspace{-0.8cm}
   \includegraphics[angle=90,width=8.7cm]{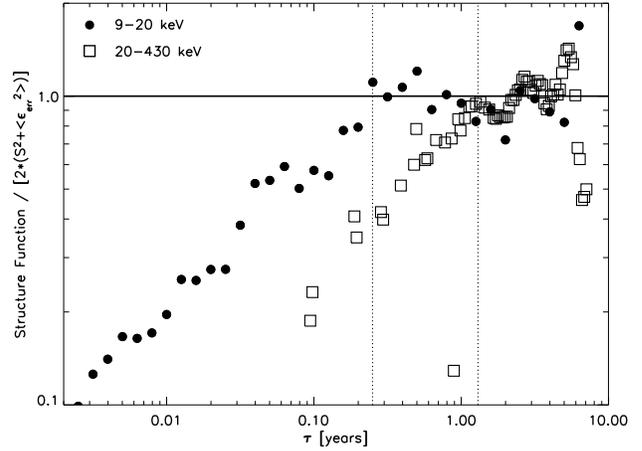}
      \caption{X-ray structure functions in the 9--20 and 20--430 keV range, based on \rxte\ (circles) and BATSE (squares) data, respectively.
      	       Both SFs have been renormalised to the expected value of the respective upper plateau (continuous line). 
	       The maximum time scales of variability are found to be around 0.22 and 1.3 years for the lower and higher energies, respectively
	       (vertical dotted lines).
               }
         \label{fig:X-raySF}
   \end{figure}

In the X-rays, a significant difference can be seen in the maximum variability time scales below and above 20 keV.
In fact, the BATSE 20--70 and 70--430 keV SFs reach the upper plateau at a time $\tau_{\rm max} = 1.3 \pm 0.4$ years,
hereas in the 4--9 and 9--20 keV bands the flattening of the SFs happens at $\tau_{\rm max} = 0.219 \pm 0.003$ years.
The difference in the measured maximum time scales can be clearly seen in Fig.~\ref{fig:X-raySF}, where the 
9--20 keV and the 20--430 keV (from combining the two BATSE data sets) SFs are compared, on time scales from 1 day to 10 years.
Even though the longer-timescale variations of the BATSE light curve occured before 1996, i.e. outside the period covered simultaneously with \rxte,
the 50 and 100 keV SFs (not containing BATSE data, but including \rxte\ data) are consistent with BATSE results, showing a smooth flattening 
in the 0.5--2 years range.

\subsection{Cross-correlation analysis}\label{croco}

We correlate all the data sets overlapping for at least 4 years and with a relatively dense sampling during this period,
concentrating on the correlations with the X-rays where this analysis is possible for the first time.
We do not discuss here optical-UV correlations, already presented by \citet{paltani98a} and for which we obtain similar results.

Two methods can be used to estimate the correlation function: the interpolated correlation analysis (ICF) where both signals are
independently interpolated,
and the discrete correlation analysis (DCF), where the correlation function is calculated in several time bins. 
Both methods have strengths and weaknesses depending on the data and their sampling \citep{white94}.
We apply both the ICF and the DCF methods (using the improvements
proposed by \citealt{gaskell87} and \citealt{white94}) and use the consistency between ICF and DCF as a criterion for 
the robustness of the correlations.
In addition, we estimate the uncertainties on the correlation functions using a bootstrap procedure.
For better clarity, only for one DCF (top left panel in Fig.~\ref{fig:croco}) and one ICF (centre left panel in Fig.~\ref{fig:croco})
we show the error bars in the Figures.
The time lags are corrected for the effect of redshift.
We discuss separately in the following the correlation of the medium ($\leq$ 20 keV, mainly \rxte\ data) and hard (20--70 keV, BATSE data) 
X-ray curves with the other wavelengths.
\subsubsection{Significant correlations}
\begin{figure}
   \centering
   \includegraphics[angle=0,width=4.2cm]{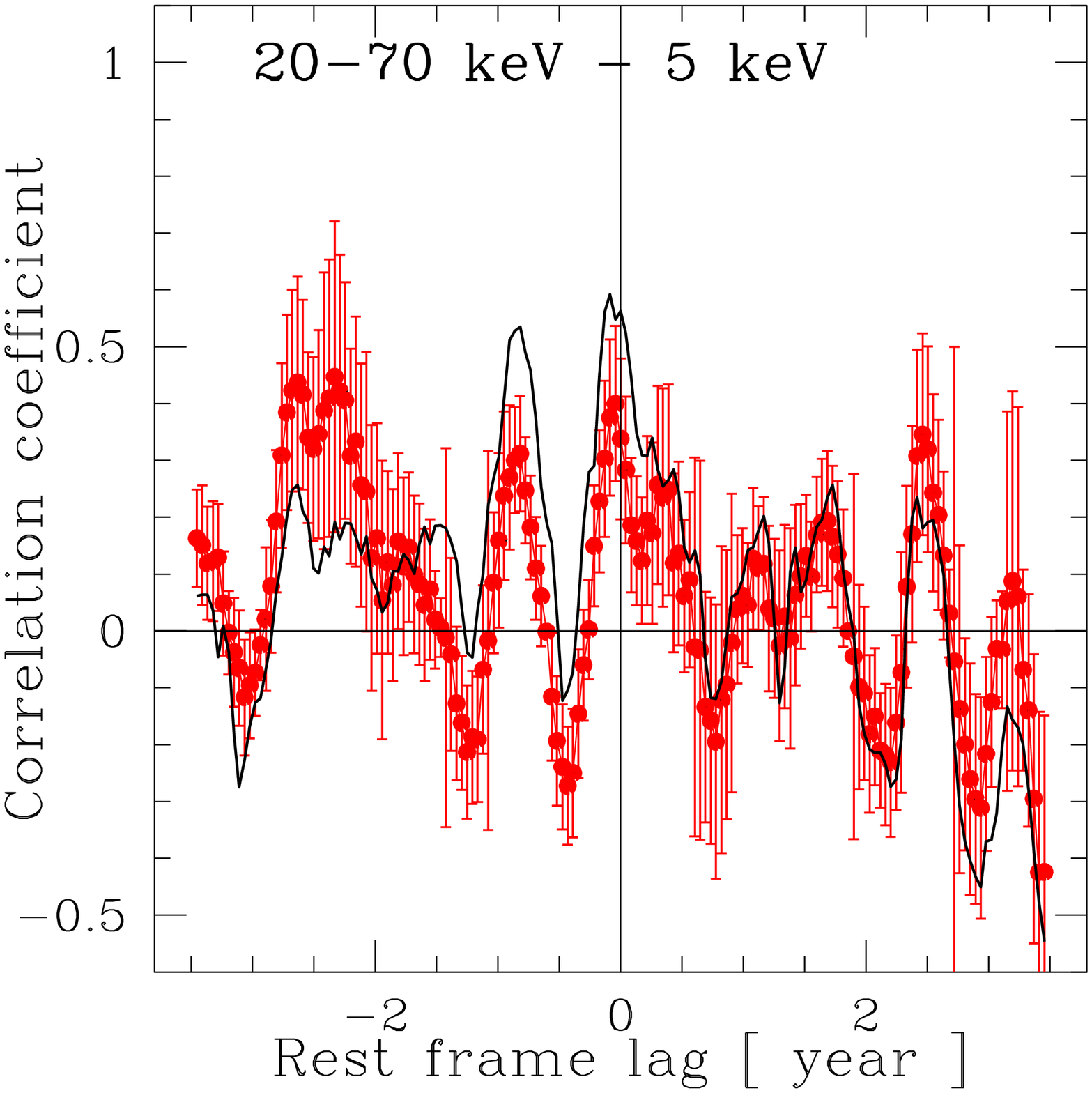}
   \includegraphics[angle=0,width=4.2cm]{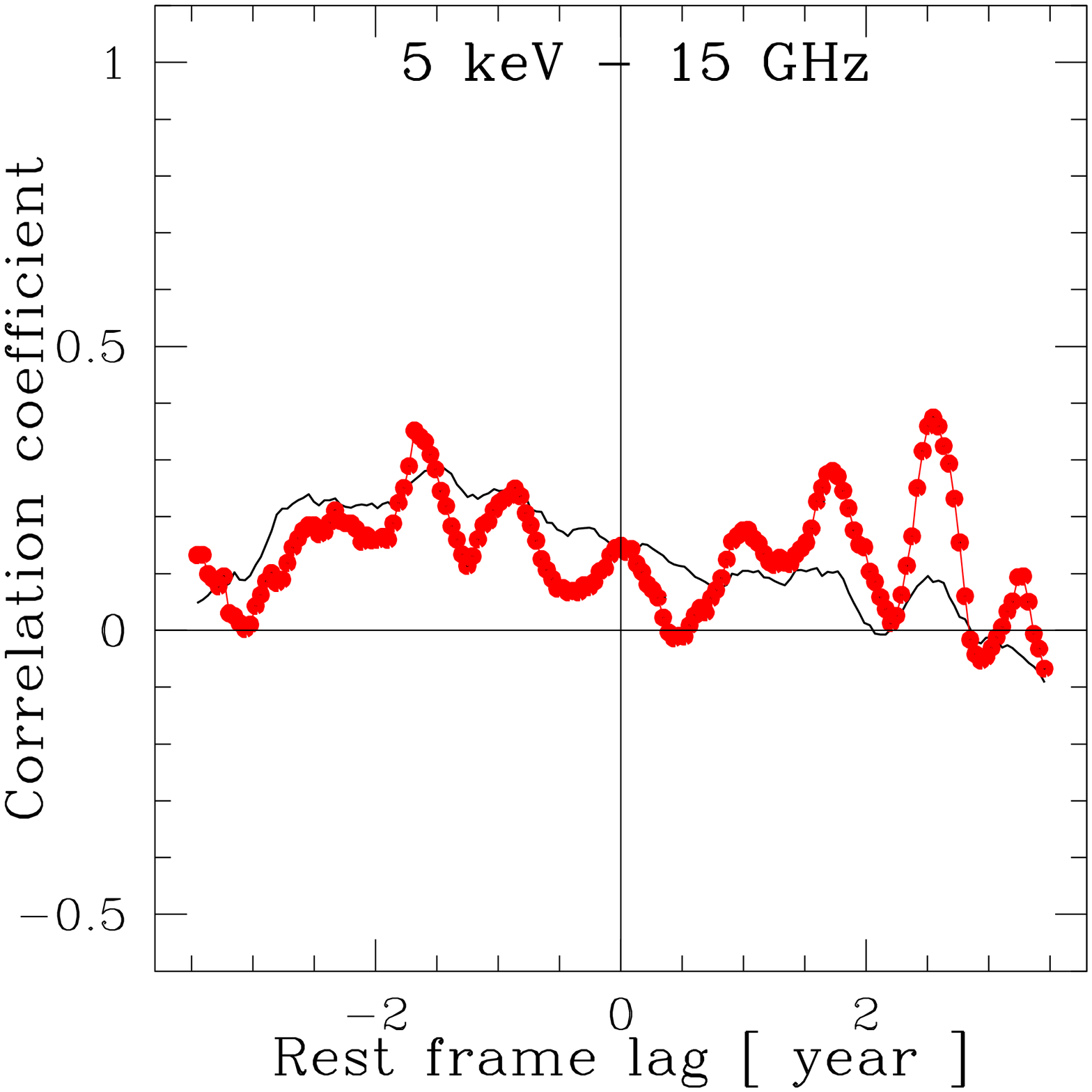}
   \includegraphics[angle=0,width=4.2cm]{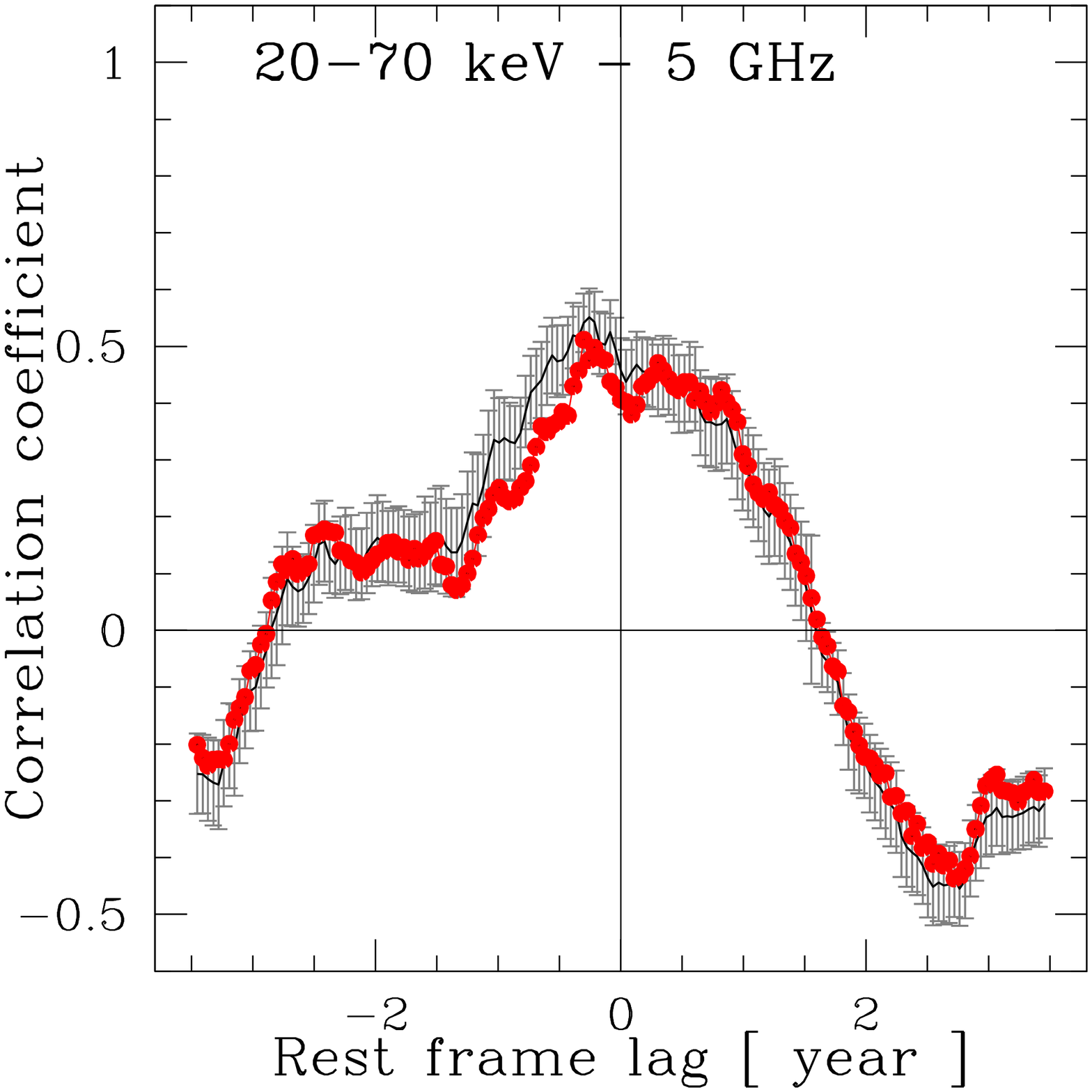}
   \includegraphics[angle=0,width=4.2cm]{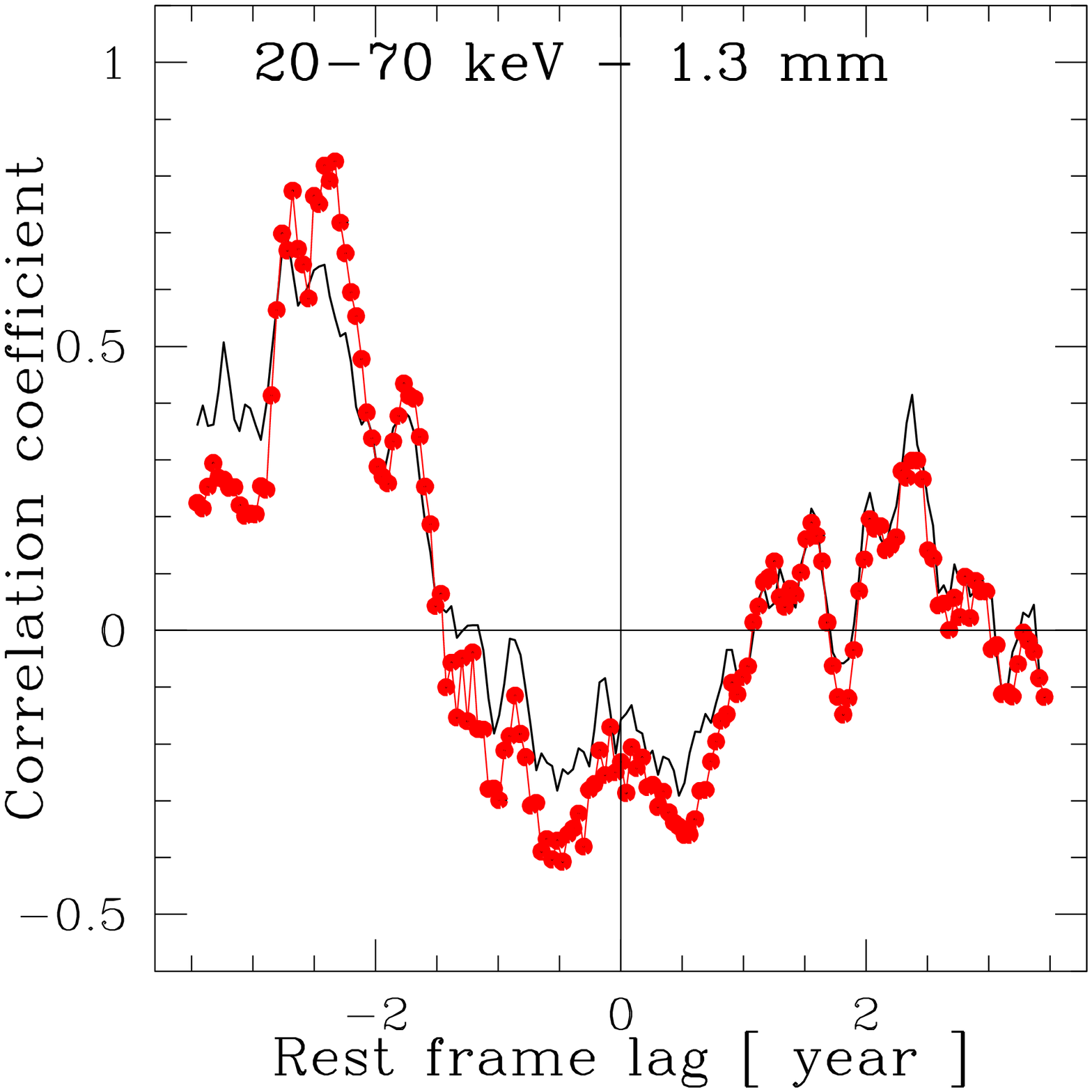}
   \includegraphics[angle=0,width=4.2cm]{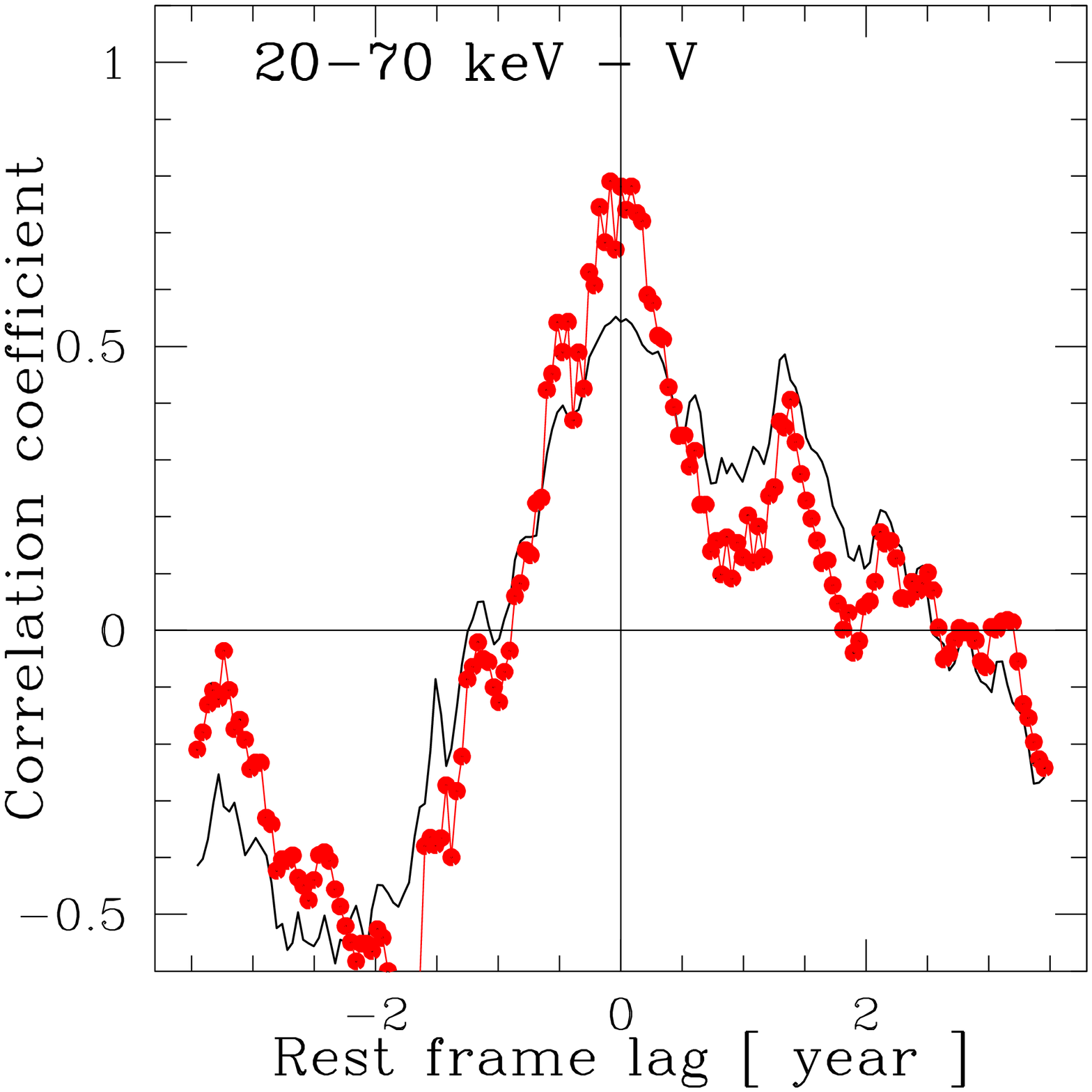}
   \includegraphics[angle=0,width=4.2cm]{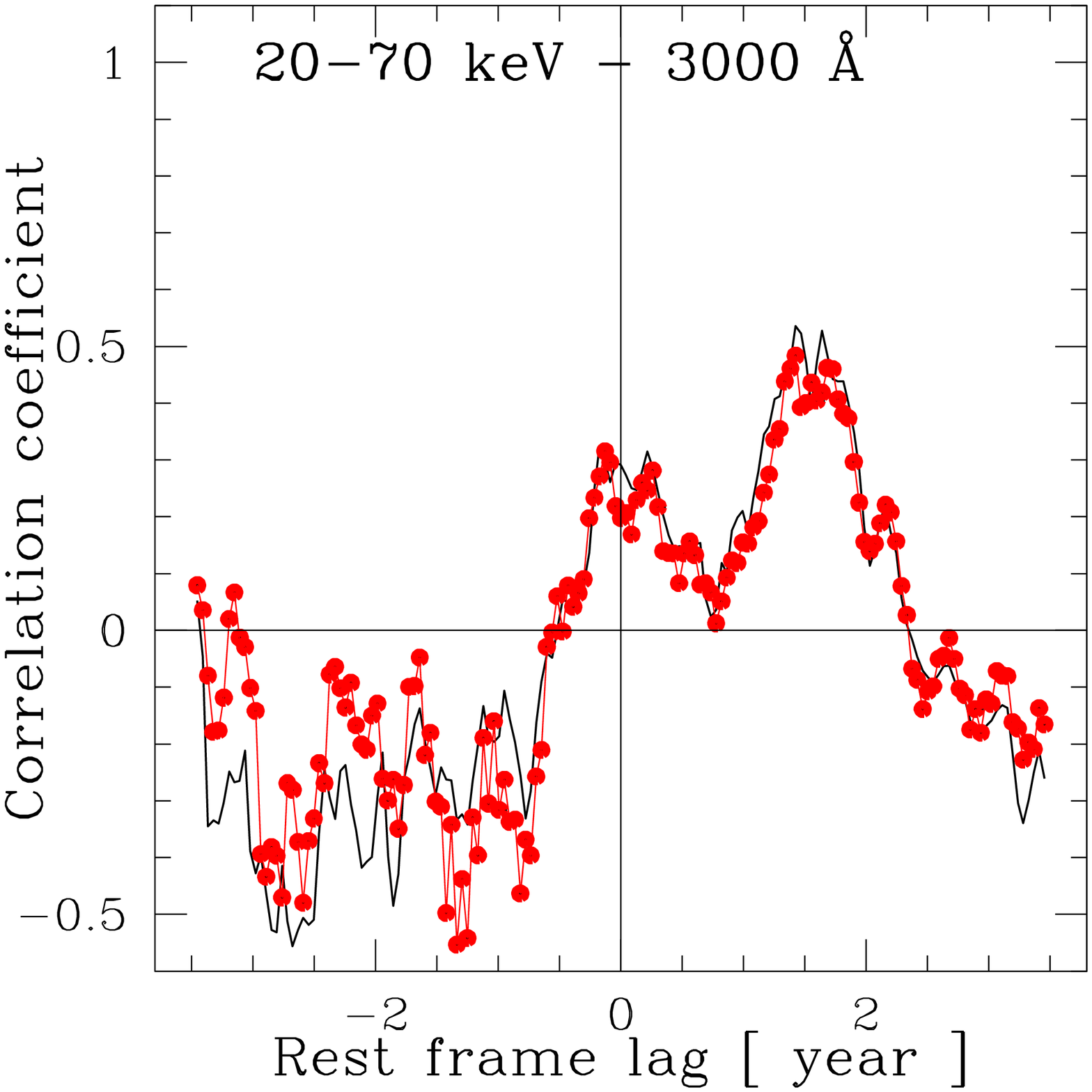}
      \caption{Cross-correlation functions as calculated with the ICF (continuous lines) and the DCF (circles) methods.
      In a correlation `$X$'--`$Y$', a positive time lag means that $Y$ is leading.
      \emph{Top left panel}: 20--70 keV  -- 5 keV, with error bars on the DCF.
      \emph{Top right panel}: 5 keV -- 15 GHz.
      \emph{Centre left panel}: 20--70 keV -- 5 GHz, with error bars on the ICF.
      \emph{Centre right panel}: 20--70 keV -- 1.3 mm.
      \emph{Bottom left panel}: 20--70 keV -- V.
      \emph{Bottom right panel}: 20--70 keV -- 3000 \AA.      
               }
         \label{fig:croco}
   \end{figure}

\begin{itemize}
\item \textit{Radio and mm:} the radio and mm light curves are very well correlated, showing peaks with correlation coefficients $R > 0.6 \pm 0.01$.
The observed delays indicate that the higher-frequency emission leads the lower-frequency one (Fig.~\ref{fig:corr_radioMM}),
in agreement with what is expected from the evolution of synchrotron flares that propagate from higher to lower energies \citep{turler99b,turler00}.
   \begin{figure}
   \centering
   \includegraphics[angle=90,width=7cm]{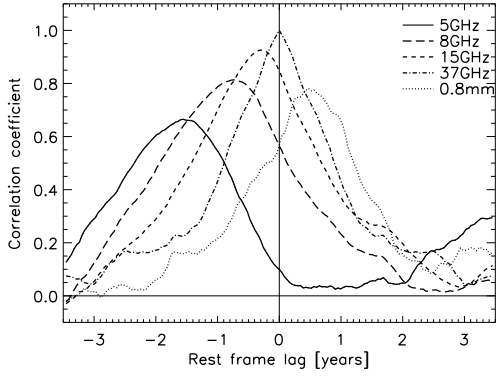}
      \caption{ICF cross-correlation functions showing the delay of the 37 GHz
      light curve on some representative radio-mm data sets at 5 GHz, 8 GHz, 15 GHz, 37 GHz (auto-correlation) and 0.8 mm.
               }
         \label{fig:corr_radioMM}
   \end{figure}
\item \textit{IR and optical-UV:} when the synchrotron flares are excluded from the data sets, a main peak is observed in the correlation between these 
light curves (Fig.~\ref{fig:corr_IR}), with correlation coefficients reaching (0.5--0.8) $\pm 0.04$ at lags between -1.2 and -1 years for the K curve 
and shifting to lower lags for  H and J (between -1 and -0.8 years for the latter). This indicates that the IR emission follows the
optical-UV one with a delay that increases with IR wavelength.
There is no clear dependence on the optical-UV frequency of the time delays and shape of the correlation function,
apart for a possible small-lag peak (at -0.3 to 0 years) for the J curve which is more prominent in the optical than in the UV correlations.
It is therefore probable that the IR correlates with both the $\mathcal{R}$ and $\mathcal{B}$ components rather than with only one of them.
   \begin{figure}
   \centering
   \includegraphics[angle=90,width=7cm]{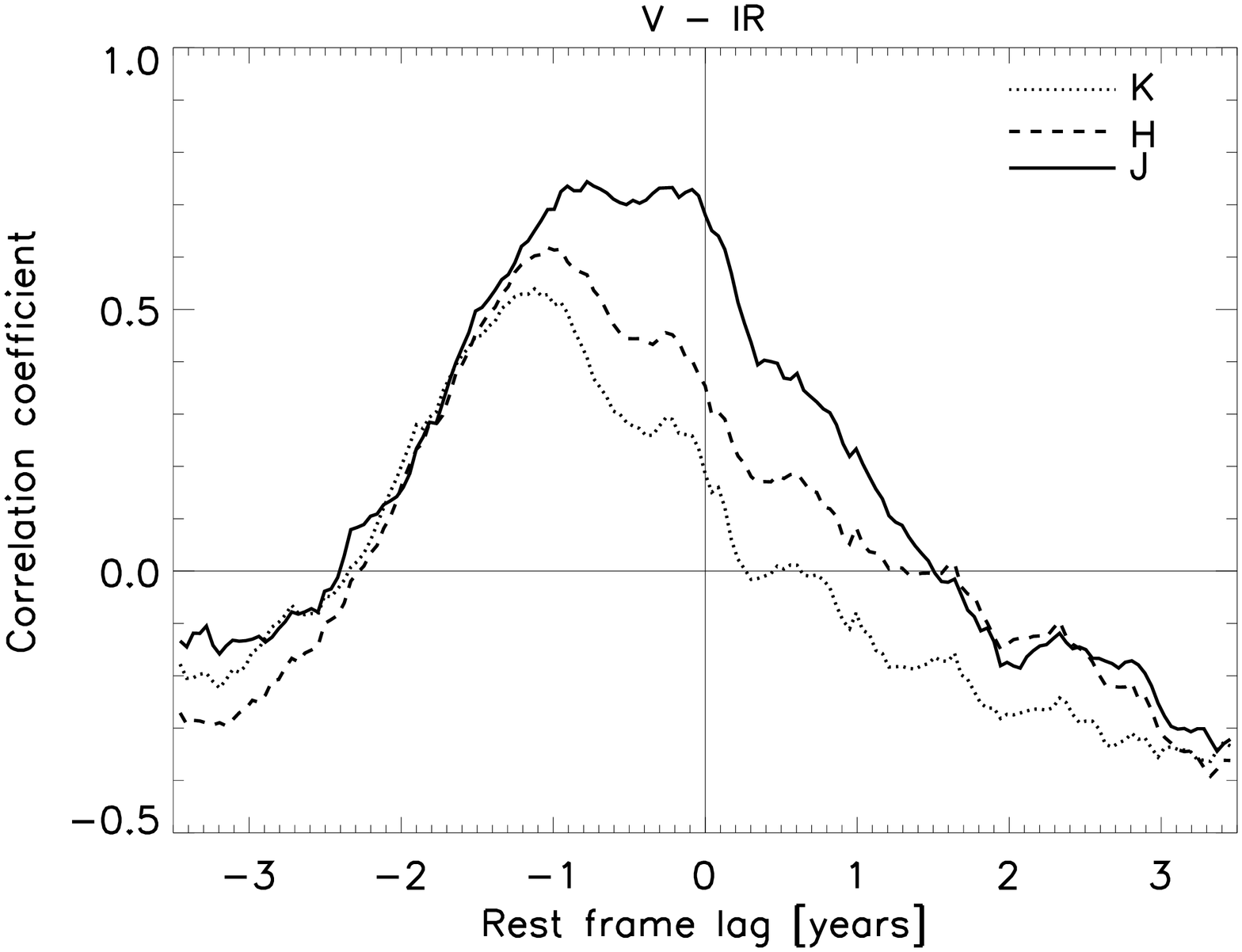}
   \includegraphics[angle=90,width=7cm]{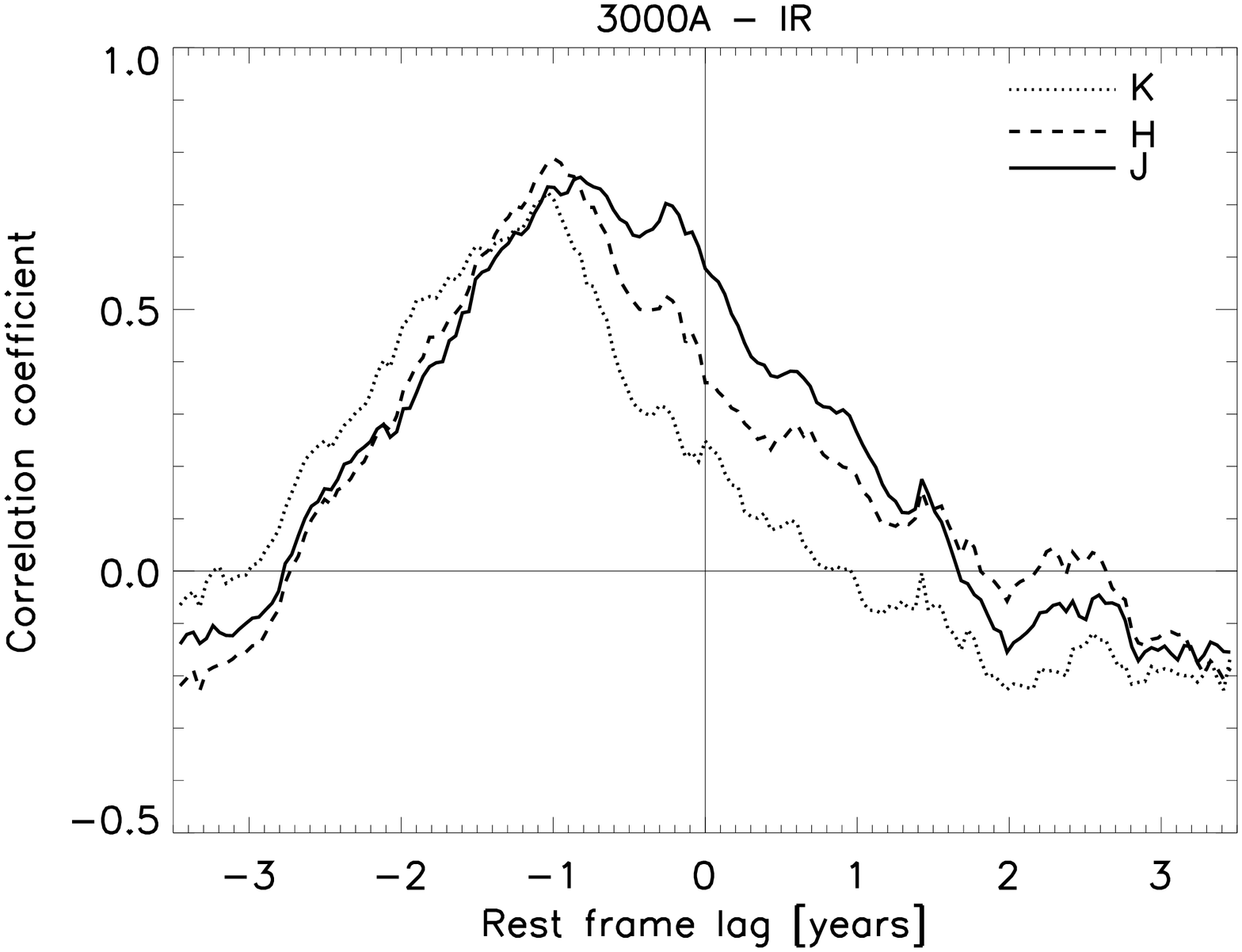}
   \includegraphics[angle=90,width=7cm]{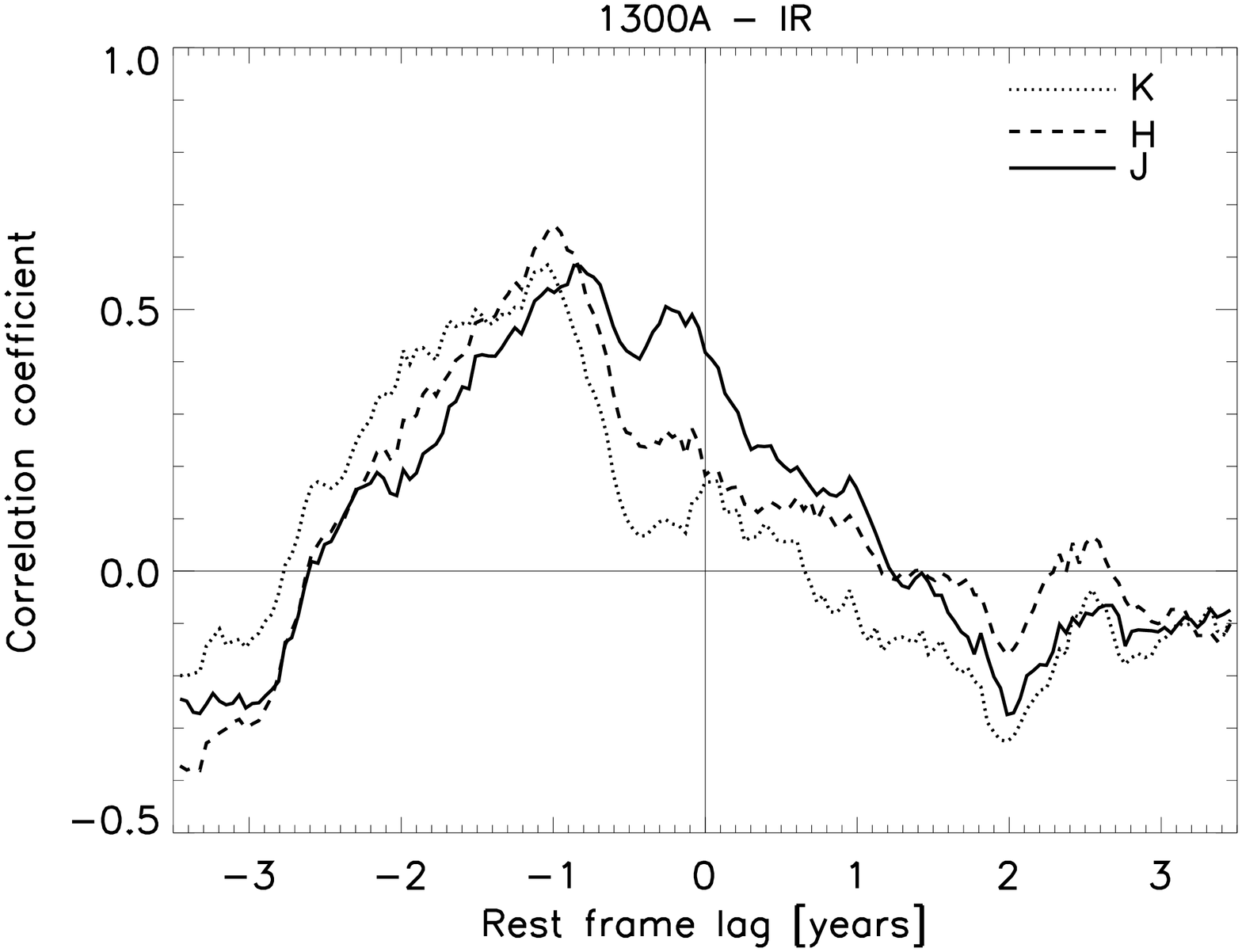}
      \caption{Cross-correlation functions (calculated with the ICF method) for the near-IR light curves K, J and H
      with some of the optical and UV data sets.
      In a correlation `$X$'--`$Y$', a positive time lag means that $Y$ is leading.
      \emph{Top panel}: V -- IR.
      \emph{Middle panel}: 3000 \AA\ -- IR.
      \emph{Bottom panel}: 1300 \AA\ -- IR.
               }
         \label{fig:corr_IR}
   \end{figure}
\item \textit{Hard X-rays and radio-mm:} there is a correlation between the 5--37 GHz light curves and the hard X-rays, with the higher radio 
frequencies leading and the X-rays correlating at about 0 lag with the 5 GHz emission (centre left panel in Fig.~\ref{fig:croco}). 
The correlation peaks reach $R = (0.5-0.6) \pm 0.06$
and are quite broad, so that a precise time lag is difficult to estimate. 
The cross-correlation plots with the mm band show a significant peak at $\sim -2.5$ years (centre right panel in Fig.~\ref{fig:croco}), which
would indicate that the X-rays are leading. 
The radio-mm correlation with the hard X-rays is due to the matching of the single, strong peak
characterising the BATSE light curve with one synchrotron flare (Fig.~\ref{fig:lc}),
i.e. the flare propagating in 1991--1993 from 37 to 5 GHz and the 1997 flare in the mm.
This strongly questions the credibility of the correlations, and therefore,
a more complete theoretical understanding is needed to know whether this correlation is due to chance occurrences or 
whether there is a connection between the hard X-ray and the jet emission and a common mechanism is responsible for both emissions.
\item \textit{Hard X-rays and optical-UV:} there are two peaks in the correlations between these bands \citep[see][]{paltani98b}, 
one at $\sim$0 lag and the other at about 1.5 years, with the blue bump emission leading the X-rays.
The peak without delay is dominant in the optical-X-ray correlation
(bottom left panel in Fig.~\ref{fig:croco}) and the peak at 1.5 years is dominant in the UV-X-ray one (bottom right panel in Fig.~\ref{fig:croco}).
This seems to indicate that the $\mathcal{R}$ component, dominant in the optical but still contributing in the UV, 
correlates at 0 lag with the X-rays, whereas the $\mathcal{B}$ component is emitted 1.5 years earlier.
These correlations, as those between the radio-mm emission and the hard X-rays, are driven by the strong peak of the BATSE light curve.
However, the less structured light curves in the optical (compared to the flaring structure of the radio-mm ones) 
and the fact that we find exactly a 0-lag correlation make the optical-hard X-ray relation more reliable. 
Since the UV-hard X-ray correlation shows a similar shape to the optical-hard X-ray one, 
the reliability of the former is likely related to that of the latter.
\end{itemize}
\subsubsection{No correlations}
\begin{figure}
   \centering
   \hspace{-1.0cm}
   \includegraphics[angle=90,width=7cm]{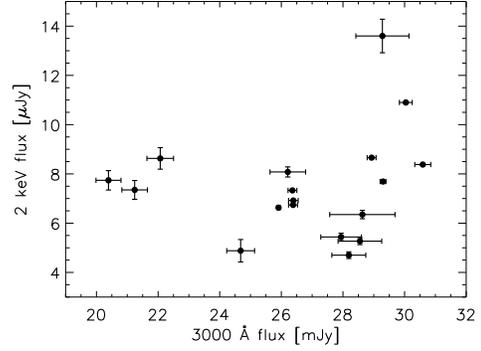}
      \caption{2 keV versus 3000 \AA\ flux obtained from quasi-simultaneous (within one day) observations.
               The Spearman-rank test results in a correlation coefficient of $R = 0.25$ and a probability for chance occurrence of $P = 0.3$,
	       indicating that there is no correlation between these two data sets.
               }
         \label{fig:corr_UVX}
   \end{figure}

\begin{itemize}
\item \textit{Medium and hard X-rays:} this correlation function shows a few peaks and repeated oscillations 
(top left panel in Fig.~\ref{fig:croco}).
Two main peaks are obtained with the ICF method at 0 and -1 year with $R \sim (0.5-0.6) \pm 0.1$, whereas in the DCF 
these peaks are at $R \sim (0.3-0.4) \pm 0.15$ and an additional peak appears at lags of --2.5 years, with $R \sim 0.45 \pm 0.2$. 
The discrepancies between the ICF and DCF and the large error bars of the DCF imply that any correlation is questionable.
\item \textit{Medium X-rays and radio-mm:} no correlation is found (top right panel in Fig.~\ref{fig:croco}), suggesting 
that the main medium-X-ray variability is not related to the jet flaring emission.
A correlation between the IR and the medium X-rays during two flares has been interpreted as evidence for an SSC origin of 
this X-ray emission \citep{mchardy07}. The lack of correlation between the medium X-rays and the radio-mm suggests that the
SSC mechanism identified by \citet{mchardy07} would rather be a secondary effect.
\item \textit{Medium X-rays and optical-UV:} as the overlap between these data sets is not sufficient for a correlation analysis on time scales of years,
we concentrate on quasi-simultaneous (within one day) optical-UV and X-ray observations.
We apply a Spearman-rank test to quantify the correlation in these subsets 
of data, containing between 6 and 50 data points. 
Apart for one significant correlation between 1 keV and 3000 \AA\ (correlation coefficient $R=0.7$ with a 
probability for chance occurrence $P=0.002$) and a marginal one between 5 keV -- B ($R=0.4$, $P=0.03$), 
no other correlation is found (see Fig.~\ref{fig:corr_UVX} for an example). 
This result questions the connection between optical-UV and X-ray emission 
which is expected in reprocessing models (see Sect.~\ref{discussion_B}).
\end{itemize}
\subsection{X-ray spectral evolution}
We study the evolution of
the X-ray photon index $\alpha$ obtained from spectra extracted within the 2--20 keV range,
to exclude the soft-excess variations and the hard-energy domain.
We concentrate our attention on the last 10-years data (1996-2006) as they represent the best-sampled period
(see \citealt{chernyakova07} for a discussion of $\alpha$ variations over 30 years).

\begin{figure}
   \centering
   \hspace{-0.8cm}
   \includegraphics[angle=90,width=9.3cm]{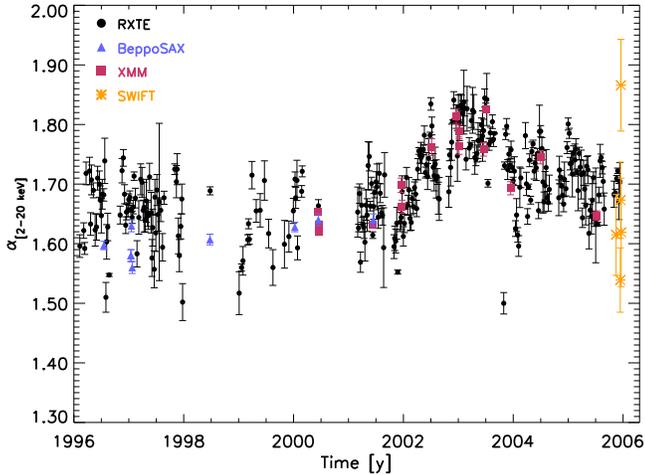}
      \caption{Evolution of the X-ray, 2--20 keV, photon index between 1996 and 2006. 
      {\it RXTE} data are rebinned to one week for better clarity.
               }
         \label{fig:gamma}
   \end{figure}

The 2--20 keV photon index has an average value of $1.71 \pm 0.07$ and exhibits significant variations of about 
$\Delta\alpha = $ 0.2 on short time scales (Fig.~\ref{fig:gamma}).
It reached a peak of $\alpha_{\rm peak} \simeq 1.84$ 
during 2003 and since then the spectrum has been hardening again.
This feature dominates the evolution of the spectral shape of the last 10 years
and it does not seem to correspond to any of the trends observed in the light curves of our database.
In fact, no correlation is found with the light curves that significantly overlap the 1996--2006 photon index curve,
i.e. the radio, 3.3 and 0.8 mm and the X-ray ones.
We investigate in more detail the relation between the 2--20 keV photon index and the 2--20 keV flux.
Depending on the time range considered, there is no correlation (e.g. 1996--2006), correlation 
(1999--2000, Spearman correlation coefficient $R$ = 0.55, $P = 3 \times 10^{-3}$) and anti-correlation 
(2001--2005, $R$ = --0.26, $P = 10^{-4}$) between X-ray photon index and flux.
These relations appear to represent random occurrences rather than a real photon index-flux connection. 
If the 1999--2000 correlation were the signature of the accretion disc or corona component, as suggested by
\citet{kataoka02}, a similar trend should be observed in 2001--2005 data when the 
jet emission reached its historical minimum \citep{turler06} and the disc/corona component should have dominated the X-ray spectrum.
Instead, an anti-correlation is observed in 2001--2005, in contradiction with this interpretation.

\section{Discussion}

\subsection{The synchrotron flaring emission}
\citet{turler99b, turler00} have extensively studied the variability behaviour of \thc\ in the radio-mm domain and
shown that it can be well reproduced with the superposition of synchrotron flares from shock waves, starting as short flares at high frequencies and
propagating to lower frequencies while evolving on longer time scales.
This scenario explains the decreasing amplitude of the variations (Fig.~\ref{fig:Fvar}), the increasing characteristic time scales (Fig.~\ref{fig:taumax}) 
and the correlations (Fig.~\ref{fig:corr_radioMM}) from the mm to the radio band.
It is likely that some flares start as very short and peaked spikes in the IR or even visible bands \citep{courvoisier88}.

\subsection{The origin of the $\mathcal{R}$ component}
The two components $\mathcal{R}$ and $\mathcal{B}$ proposed by \citet{paltani98a} to model the blue bump
show different spectral and variability properties, suggesting different origins
of the underlying emission processes.

In view of the measurement of slightly polarized emission in the optical \citep[even outside the periods of synchrotron flares,][]{dediego92}, 
one natural explanation is that $\mathcal{R}$ could be synchrotron emission not directly related to the radio-mm jet (Fig.~\ref{fig:sed_IR}).
The small amplitude of the variations observed in the $\mathcal{R}$ component (Fig.~\ref{fig:Fvar}) could be explained 
by an electron population with cooling time longer than the injection time in the emitting region.
Even though $\mathcal{R}$ is not completely constrained by the decomposition of \citet{paltani98a},
its global spectral shape is also compatible with optically thin synchrotron emission.

We investigate also the possibility that $\mathcal{R}$ origins in the broad line region (BLR). 
Considering that optical emission lines are estimated to contribute at most by 10\% of the total continuum \citep{paltani98a}
and an additional 10\% is expected from the Balmer continuum and the Fe\textsc{ii} emission \citep{courvoisier87},
the remaining $\mathcal{R}$ emission could have a thermal origin, as Bremsstrahlung radiation from the BLR.
This possibility is supported by the fact that, even though most of the flux variations of 
Ly$\alpha$ and C\textsc{iv} has been found to correlate with the photoionising UV emission, 
the long-term variability properties of these lines are more similar to those of the $\mathcal{R}$ component \citep{paltani03}.
From photo-ionisation models \citep{robson96}, we calculate a BLR mass of 160 $\rm M_\odot$, using an average H$\alpha$ flux of 
$6 \times 10^{-12} \, \rm erg \, cm^{-2} \, s^{-1}$ \citep{kaspi00} and assuming uniform electron density of the BLR with
a canonical value of $10^{10} \rm \, cm^{-3}$. If the BLR gas is clumpy, the BLR mass is expected to be less
than an order of magnitude larger.
A mass of 160 $\rm M_\odot$ at 3000 K, as suggested by the spectral shape,
results in a Bremsstrahlung luminosity of $L_{\rm Bremss} = 1.7 \times 10^{44} \rm \, erg/s$, which is only 1\% of the total 
luminosity observed in the $\mathcal{R}$ component
($4 \times 10^{46} \, \rm erg/s$), ruling out a BLR Bremsstrahlung origin for $\mathcal{R}$.

\subsection{$\mathcal{B}$ and X-rays relation}\label{discussion_B}
A short-lag correlation between the UV and X-ray ($\leq$ 10 keV) emissions is expected if reprocessing 
is the main process producing the UV radiation.
In the past, such a correlation has been found \citep{chernyakova07} or not \citep{walter92b} 
on smaller data sets, depending on the time period.
We obtain here that there is no short-lag correlation between the light curves on these two bands, indicating
that, if reprocessing is the driving process (as suggested by the UV spectral and variability properties, \citealt{paltani98a}), 
somehow the X-ray radiation interacting with the UV source is not the same as that observed from Earth.
This could happen if the X-ray emission is anisotropic \citep{uttley05} or if the X-ray source heating the disc is positioned 
very close to the black hole.
In this case, only a small part of the X-rays could escape, whereas most of them would be gravitationally 
bent by the black hole and illuminate the accretion disc.
This possibility  based on the light bending model of \citet{miniutti04} needs proper simulations in order 
to prove that this geometry would actually reproduce the observed emission. 

Several studies on Seyfert galaxies have given conflicting results on the optical/UV-X-ray relation, 
showing correlation \citep[\object{NGC 5548},][]{uttley03} or no correlation 
\citep[\object{NGC 3516},][]{maoz02}, sometimes depending on the observed time scale \citep[\object{NGC 4051},][]{shemmer03}, and with different 
variability amplitude in the optical/UV compared to the X-rays (see \citealt{uttley05} for a review). 
There are however several reasons for which physically connected blue bump and X-ray emissions would not be correlated.
For example if the X-rays are due to Comptonisation of the UV photons, the UV-X-ray flux relation could depend on a third parameter,
e.g. the X-ray spectral index \citep{nandra01}.
Yet, even though such a relation was detected in \thc\ in the past by \citet{walter92b}, 
when more recent data are added no correlation of these three parameters is found \citep{chernyakova07}.
Among other possibilities, the feedback between reprocessing and Comptonisation or the geometry and the size of the 
X-ray source or of the reprocessing medium could wash out the variations of the observed emission \citep{nandra01}.

An alternative scenario for the production of UV radiation involves matter accreted in clumps, rather than through an accretion disc \citep{courvoisier05}.
The interaction of these clumps at $\sim 100$ Schwarzschild radii generates optically thick shocks producing the UV emission, whereas
optically thin shocks closer to the black hole give origin to the X-rays. In this model, both a short-lag and a longer-lag correlation
are expected. The latter correlation has been found in \thc\ but between the hard X-rays and the UV. 
The application of this model to the case of \thc\ will be the subject of a forthcoming paper.
\subsection{The near-IR emission}
The amplitude of the variations (Fig.~\ref{fig:Fvar}) and the shape of the near-IR to UV spectrum (Fig.~\ref{fig:sed_IR})
suggest that $\mathcal{R}$ could extend to near-IR wavelengths and be responsible for the variability in this band, 
synchrotron flares excepted.
An additional dust contribution has been proposed in the past \citep{robson83,turler06}
and is confirmed here by the need of a thermal component to model the average near-IR spectrum.
As the extrapolation of the mm emission is higher than the near-IR flux, a cut-off is required in the high frequency
synchrotron emission at $\sim 5 \times 10^{13}$ Hz,
and therefore $\mathcal{R}$ and dust emission are likely the dominant contributors in the near-IR.

The interplay between the components is observable in the correlations between the near-IR and the optical-UV emission
(Fig.~\ref{fig:corr_IR}).
In the most likely scenario, the dust is heated up by the UV/$\mathcal{B}$ continuum.
The complexity of the dust component appears in the 0.8--1.2 y delays found in the correlations. These could represent the travelling time of the 
$\mathcal{B}$ photons from the UV source to the dust location. The delays increase with increasing IR wavelength, which is 
expected from an extended dust region with a distribution of temperatures.
A similar correlation between near-IR and optical/UV has been observed in a number of Seyfert 1 galaxies. The time lag of the K 
with the V band seems to be proportional to the square root of the optical luminosity \citep[][and references therein]{suganuma06} and is 
believed to represent an estimate of the inner radius of the dust torus in these objects. 
Even though \thc\ is 100 times more luminous than the Seyfert 1 \object{Fairall 9}, their time lags are similar
\citep{clavel89}, indicating that \thc\ does not follow the lag-luminosity relation of Seyfert 1 galaxies.

The correlation of the J light curve with the optical-UV shows a secondary peak at short lags, which is higher in the optical 
than in the UV band (Fig.~\ref{fig:corr_IR}). The same effect, even though much weaker, is still present in the correlation of H with the optical-UV. 
This secondary peak could be the signature of the $\mathcal{R}$ component, dominant in the optical and responsible for most of the observed emission
also in the J band.
The same effect has been reported for Fairall~9, for which the J band was found to be correlated without lag to the optical-UV curves, 
whereas K and L were correlated with delays of $\simeq$ 1 year \citep{clavel89}.
This correlation represents an additional evidence that $\mathcal{R}$ contributes to the near-IR emission of \thc.
\subsection{The origin of the X-ray emission}\label{origin_X}
In the X-rays, the maximum variability time scales are significantly longer above 20 keV than below and the
amplitude of the variations is increasing steeply with energy above 20 keV, whereas at lower
X-ray energies the increase is rather shallow.
In addition, no convincing correlation is detected between the light curves below and above 20 keV.

The difference between medium and hard X-rays imply the presence of two
varying parameters, which can be interpreted either as two distinct physical
components or as a single component with varying spectral shape.
Both descriptions are equivalent, they may, however, lead to different scenarios. 

In the two-components scenario, the high-energy emission of \thc\
could be due to the superposition of a Seyfert-like and a blazar-like component, as suggested by e.g. \citet{grandi04}.
Comptonisation of thermal plasma for example in an accretion disc would produce X-ray emission 
similar to that observed in Seyfert galaxies \citep{grandi04}.
The occasional detection of a iron K$\alpha$ emission line \citep[e.g.][]{kataoka02,grandi04,chernyakova07}
confirms the similarity between \thc\ and Seyfert emission in the medium X-rays.
Above 10--20 keV, inverse Compton on a population of relativistic electrons from a non-thermal plasma would produce the observed 
high-energy emission, as suggested by the presence of a break in the spectrum around 1 MeV \citep{johnson95,lichti95}. 

In the single-component scenario, the 2 to 500 keV spectrum could be produced, for example, by a single power law
pivoting at low X-ray energies. 
Then the variations at medium X-rays would be dominated by the variation in the low energy flux of this component,
whereas the variations at hard X-rays would rather be due to the variation of the spectral index.
The long time-scale variations of the spectral index (Fig.~\ref{fig:gamma}) and of the hard X-ray emission (Fig.~\ref{fig:X-raySF}) could support this
decomposition.

The lack of short-lag correlation between the mm and the X-ray radiation indicates that there is no clear 
connection between the X-rays and the radio-mm flaring emission from the jet.
Compton losses are thought to be the dominant electron cooling process in the first phase of the evolution of the shocks in the jet 
\citep{marscher85}. Therefore, if this process were producing the bulk of the X-ray emission,
the mm flares would follow the X-ray ones with short delays, which is not observed.
In addition, the hard X-ray peak in the 20--70 keV light curve is broad (Fig.~\ref{fig:lc}), also not expected 
if it represented the early phase of the shock flaring emission.
The lack of correlation between the medium X-ray and the mm suggests that the
simultaneous IR and medium X-ray emission detected by \citet{mchardy07} and interpreted as due to SSC from the jet must therefore be the result of 
a secondary process. This shows that some synchrotron events can be so energetic that they dominate the X-ray emission even up to 20 keV
for short time periods. 

A correlation without delay is observed between the hard X-rays and the $\mathcal{R}$ component. 
If $\mathcal{R}$ is synchrotron emission, as suggested here, the hard X-rays could be produced by the same electron population emitting 
the $\mathcal{R}$/synchrotron component.
An alternative to this synchrotron self-Compton mechanism (SSC) is an external Compton process (EC) suggested by the 1.5 y delay observed
between the $\mathcal{B}$ emission and the hard X-rays.
If one assumes that $\mathcal{B}$ provides the seed photons for the Compton processes 
producing hard X-rays, this 1.5-year delay could indicate the distance covered by the photons from the UV to the X-ray source.


\section{Summary and conclusions}

We have presented an update of the multiwavelength \thc\ database with the last 10 years of data and performed an analysis
of the variability properties of this object with different methods.

In order to explain the significantly different timing properties of the X-ray emission below and above $\sim$ 20 keV, 
either two separate components (possibly a Seyfert-like and a blazar-like one) or at least two parameters with distinct variability properties
are needed.
The 5--500 keV emission does not seem to be related to the mm-jet, questioning the usually assumed common origin for the electrons emitting 
the radio-mm and the hard X-ray radiations.
On the other hand, the electron population at the origin of the dominant hard X-ray emission could be the same as that responsible for the optical emission,
which is likely to originate from synchrotron mechanism. If the Comptonised photons are the UV ones, the 1.5-year delay
between the UV and the hard X-ray emission could be an indication of the distance between the X-ray emitting region and the seed photon source.

The blue-bump emission does not seem to be related in an obvious way to the X-ray emission below 20 keV. If there is a close interplay between 
the blue bump and the X-rays, as expected in Comptonisation and reprocessing models, then somehow the medium-X-ray emission we observe is not the same
as that interacting with the blue bump, maybe due to some particular geometry of the emitting region or anisotropy of the X-ray radiation,
possibly because of light bending effects.

The near-IR spectral and timing properties are well explained by the combined contribution of emission from dust illuminated by a UV source 
at $\lesssim$ 1 light year and the extension of the optical/synchrotron component. 

The variability properties of the radio-mm emission are well understood in the frame of synchrotron flares from shock waves 
propagating along the jet, with spectral and timing characteristics evolving with time. Assuming that the maximum variability 
time scales measured in this band represent an upper limit to the electron cooling time via synchrotron emission, we obtain an 
estimate of the magnetic field of $B \gtrsim 0.09$ Gauss.

The complexity of \thc\ emission (and of AGN in general) is, at least in part, due to the presence of many
emission components of different nature but similar luminosities, probably originating in different locations within 
the AGN. In addition, in several energy domains the observed radiation is produced by the superposition of 
two or more of these different components.
This highlights the importance of long-term multiwavelength studies for a better understanding of the physics 
of these objects.
An easy access to all observational data is therefore crucial to be able to use them even a long time after the observations were performed.

\begin{acknowledgements}
We thank M. Dahlem, R. J. Sault and the ATCA observatory staff for providing the mm data and 
an estimate on their accuracy.
We also thank M. Gurwell for providing mm/sub-mm data from the SMA Calibrator Database
and J. Carpenter for flux measurements in the mm band from the OVRO database.
AL and MT acknowledge support for Mets\"ahovi observing projects from the Academy of Finland.
The long-term UMRAO monitoring program has been supported by a series of grants from the NSF, most
recently AST-0607523, and by funds from  the University of Michigan.
This work includes data from the OMC Archive at LAEFF, pre-processed by ISDC.
\end{acknowledgements}


\bibliographystyle{aa}
\bibliography{biblio}

%
\setlength{\LTcapwidth}{7in}
\clearpage
\onecolumn
\addtocounter{table}{-2}
\begin{longtable}{l c c c c c c c c}
\caption{\label{table:var} Radio to $\gamma$-rays 3C 273 light curves used for the variability analysis.
Col.1: light curve name. 
Epoch: the epoch between the first and last observation.
$N$: number of data points used for the variability analyses. 
$\overline{\nu}$: average frequency.
$\overline{F_{\nu}}$ and $\sigma_{\nu}$: average flux and its standard deviation in Jy.
$\overline{\nu F_{\nu}}$: average intensity in units of $\rm 10^{-11} \, erg \, cm^{-2} \, s^{-1}$.
$F_{\rm var}$: variability amplitude and its uncertainty from bootstrap simulations.
$\tau_{\rm max}$: maximum variability time scale from structure function analysis and its uncertainty from bootstrap simulations.
`$\ldots$'  indicates those light curves not used for the particular analysis due to different reasons,
for instance an insufficient number of data points.
(wf): data sets from which the synchrotron flares have been excluded. $^{(a)}$ flux in counts/s.
$^{(b)}$ flux in $10^{-10} \, \rm erg \, cm^{-2} \, s^{-1}$.
} \\
\hline\hline
\noalign{\smallskip}
Light Curve & Epoch & $N$ & $\overline{\nu}$ & $\overline{F_{\nu}}$ & $\sigma_{\nu}$ & $\overline{\nu F_{\nu}}$ & $F_{\rm var}$ & $\tau_{\rm max}$  \\ 
            &   & &       Hz       & Jy                 & Jy           & $\rm 10^{-11} \, erg \, cm^{-2} \, s^{-1}$ &              & yr \\ 
\noalign{\smallskip}
\hline
\noalign{\smallskip}
\endfirsthead
\caption{continued.}\\
\hline\hline
\noalign{\smallskip}
Light Curve & Epoch & $N$ & $\overline{\nu}$ & $\overline{F_{\nu}}$ & $\sigma_{\nu}$ & $\overline{\nu F_{\nu}}$ & $F_{\rm var}$ & $\tau_{\rm max}$  \\ 
            &   & &       Hz       & Jy                 & Jy           & $\rm 10^{-11} \, erg \, cm^{-2} \, s^{-1}$ &              & yr \\ 
\noalign{\smallskip}
\hline
\noalign{\smallskip}
\endhead
\hline
\endfoot
         2.5 GHz  & 1979--1995&  122  &  $ 2.61 \, 10^{9} $  &   42.5  &   1.9  & 0.11 &  
0.041 $\pm$ 0.003  &   $\ldots$   \\
           5 GHz  & 1967--2006 &  869  &  $ 4.93 \, 10^{9} $  &   38.0  &   3.6  & 0.19 & 
0.094 $\pm$ 0.003  &    $\ldots$    \\
           8 GHz  & 1963--2006 & 1568  &  $ 8.00 \, 10^{9} $  &   36.4  &   5.7  & 0.29 & 
0.157 $\pm$ 0.002  &   $4.4 {+0.3 \atop -0.3}$ \\
          10 GHz  & 1965--1994 &  293  &  $ 9.07 \, 10^{9} $  &   44.3  &   7.5  & 0.40 & 
0.167 $\pm$ 0.006  &    $\ldots$    \\
          15 GHz  &  1963--2006 & 1286  &  $ 1.46 \, 10^{10} $  &   32.5  &   9.6  & 0.48 & 
0.294 $\pm$ 0.005  &   $4.07 {+1.3 \atop -0.5}$  \\
          22 GHz  &  1976--2004 & 1099  &  $ 2.23 \, 10^{10} $  &   32.0  &   9.2  & 0.71  & 
0.285 $\pm$ 0.005  &   $1.6 {+0.4 \atop -0.1}$  \\
          37 GHz  &  1970--2006 & 1259  &  $ 3.69 \, 10^{10} $  &   25.0  &   9.3  & 0.92  &
0.373 $\pm$ 0.007  &   $1.23 {+0.19 \atop -0.03}$  \\
          3.3 mm  &  1965--2006 & 1548  &  $ 9.30 \, 10^{10} $  &   20.1  &  13.0  & 1.9  &
0.63 $\pm$ 0.01  &    $\ldots$    \\
            2 mm  &  1981-2004 & 204  &  $ 1.50 \, 10^{11} $  &   17.8  &   6.3  & 2.7  &
0.34 $\pm$ 0.02  &   $0.4 {+0.8 \atop -0.1}$  \\
          1.3 mm  &  1981--2007 & 545  &  $ 2.30 \, 10^{11} $  &   11.8  &   5.4  & 2.7  &
0.45 $\pm$ 0.02  &   $0.2 {+2.3 \atop -0.07}$  \\
          1.1 mm  &  1973--2007 & 322  &  $ 2.76 \, 10^{11} $  &   13.5  &   6.4  & 3.7  &
0.46 $\pm$ 0.06  &    $\ldots$    \\
          0.8 mm  &  1981--2007 & 496  &  $ 3.62 \, 10^{11} $  &    8.4  &   4.5  & 3.0  &
0.53 $\pm$ 0.03  &    $\gtrsim 13$    \\
         0.45 mm  &  1982--1996 &  79  &  $ 6.72 \, 10^{11} $  &    6.1  &   3.8  & 4.1  &
0.58 $\pm$ 0.06  &   $\ldots$   \\
         0.35 mm  &  1983--1995 &  17  &  $ 8.57 \, 10^{11} $  &    7.7  &   6.0  & 6.6  &
0.73 $\pm$ 0.10  &   $\ldots$   \\
              Q  &   1971--2004 & 17  &  $ 1.44 \, 10^{13} $  &   0.80  &   0.31  & 11.5  &
0.35 $\pm$ 0.07  &   $\ldots$   \\
              N  &   1970--2004 & 75  &  $ 2.86 \, 10^{13} $  &   0.35  &   0.09  & 10.0  &
0.23 $\pm$ 0.02  &   $\ldots$   \\
         N (wf)  &   1970--2004 & 54  &  $ 2.85 \, 10^{13} $  &   0.31  &   0.06  & 8.8  &
0.14 $\pm$ 0.04  &   $\ldots$   \\
              M  &   1983--1993 & 33  &  $ 6.33 \, 10^{13} $  &   0.27  &   0.12  & 17.1 &
0.34 $\pm$ 0.06  &   $\ldots$   \\
         M (wf)  &   1986--1983 & 21  &  $ 6.31 \, 10^{13} $  &   0.20  &   0.03  & 12.6  &
      $\ldots$   &   $\ldots$   \\
              L  &   1969--2004 & 165  &  $ 8.09 \, 10^{13} $  &   0.17  &   0.03  & 13.8  &
0.15 $\pm$ 0.02  &   $\ldots$   \\
         L (wf)  &   1969--2004 & 126  &  $ 8.09 \, 10^{13} $  &   0.16  &   0.01  & 13.0  &
0.04 $\pm$ 0.01  &   $\ldots$   \\
              K  &   1967--2004 & 350  &  $ 1.36 \, 10^{14} $  &  $ 8.5 \, 10^{-2} $  &    
$ 1.1 \, 10^{-2} $  & 11.6  & 0.120 $\pm$ 0.009  &    $\ldots$    \\
         K (wf)  &   1967--2004 & 246  &  $ 1.35 \, 10^{14} $  &  $ 8.1 \, 10^{-2} $  &  
$ 6.2 \, 10^{-3} $  & 10.9  & 0.059 $\pm$ 0.005  &    $\ldots$    \\
              H  &   1967--2004 & 317  &  $ 1.82 \, 10^{14} $  &  $ 4.8 \, 10^{-2} $  &  
$ 7.9 \, 10^{-3} $  & 8.7 & 0.16 $\pm$ 0.01  &    $\ldots$    \\
         H (wf)  &   1967--2004 & 247  &  $ 1.82 \, 10^{14} $  &  $4.6 \, 10^{-2} $  &  
$ 4.2 \, 10^{-3} $  & 8.4 & 0.076 $\pm$ 0.005  &    $\ldots$    \\
              J  &   1976--2004 & 280  &  $ 2.40 \, 10^{14} $  &  $3.6 \, 10^{-2} $  &  
$ 6.0 \, 10^{-3} $  & 8.6 & 0.16 $\pm$ 0.02  &    $\ldots$    \\
         J (wf)  &   1976--2004 & 218  &  $ 2.40 \, 10^{14} $  &  $3.4 \, 10^{-2} $  &  
$ 3.8 \, 10^{-3} $  & 8.2 & 0.102 $\pm$ 0.005  &    $\ldots$    \\
              G  &   1985--2003 & 438  &  $ 5.17 \, 10^{14} $  &  $ 3.2 \, 10^{-2} $  &  
$ 3.4 \, 10^{-3} $  & 16.5 & 0.106 $\pm$ 0.004  &    $\ldots$    \\
         G (wf)  &   1985--2003 & 396  &  $ 5.17 \, 10^{14} $  &  $ 3.2 \, 10^{-2} $  &  
$ 3.2 \, 10^{-3} $  &  16.5 & 0.101 $\pm$ 0.003  &    $\ldots$    \\
              V  &   1968--2005 & 730  &  $ 5.50 \, 10^{14} $  &  $ 2.9 \, 10^{-2} $  &  
$ 3.4 \, 10^{-3} $  & 16.0 & 0.113 $\pm$ 0.004  &   $3.9 {+4.6 \atop -3.3}$  \\
         V (wf)  &   1968--2005 & 675  &  $ 5.50 \, 10^{14} $  &  $ 2.9 \, 10^{-2} $  &  
$ 3.2 \, 10^{-3} $  & 16.0 & 0.108 $\pm$ 0.003  &    $\ldots$    \\
             V1  &   1985--2003 & 438  &  $ 5.56 \, 10^{14} $  &  $ 3.0 \, 10^{-2} $  &  
$ 3.3 \, 10^{-3} $  & 16.7 & 0.110 $\pm$ 0.003  &    $\gtrsim 9$    \\
        V1 (wf)  &   1985--2003 & 396  &  $ 5.56 \, 10^{14} $  &  $ 2.9 \, 10^{-2} $  &  
$ 3.2 \, 10^{-3} $  & 16.1 & 0.109 $\pm$ 0.003  &    $\ldots$    \\
             B2  &   1985--2003 & 438  &  $ 6.71 \, 10^{14} $  &  $ 2.7 \, 10^{-2} $  &  
$ 3.3 \, 10^{-3} $  & 18.1 & 0.122 $\pm$ 0.003  &    $\gtrsim 9$    \\
        B2 (wf)  &   1985--2003 & 396  &  $ 6.71 \, 10^{14} $  &  $ 2.7 \, 10^{-2} $  &  
$ 3.4 \, 10^{-3} $  & 18.1  & 0.126 $\pm$ 0.004  &    $\ldots$    \\
              B  &   1968--2005 & 755  &  $ 6.91 \, 10^{14} $  &  $ 2.7 \, 10^{-2} $  &  
$ 3.3 \, 10^{-3} $  & 18.7 & 0.120 $\pm$ 0.003  &   $3.9 {+0.6 \atop -0.5}$  \\
         B (wf)  &   1968--2005 & 690  &  $ 6.91 \, 10^{14} $  &  $ 2.7 \, 10^{-2} $  &  
$ 3.3 \, 10^{-3} $  & 18.7  & 0.122 $\pm$ 0.003  &    $\ldots$    \\
             B1  &   1985--2003 & 438  &  $ 7.49 \, 10^{14} $  &  $ 2.8 \, 10^{-2} $  &  
$ 3.5 \, 10^{-3} $  & 21.0  & 0.125 $\pm$ 0.004  &    $\gtrsim 9$    \\
        B1 (wf)  &   1985--2003 & 396  &  $ 7.49 \, 10^{14} $  &  $ 2.8 \, 10^{-2} $  &  
$ 3.6 \, 10^{-3} $  & 21.0 & 0.131 $\pm$ 0.004  &    $\ldots$    \\
              U  &   1968--2005 & 680  &  $ 8.59 \, 10^{14} $  &  $ 2.7 \, 10^{-2} $  &  
$ 3.6 \, 10^{-3} $  & 23.2 & 0.132 $\pm$ 0.004  &   $3.9 {+0.6 \atop -0.5}$  \\
         U (wf)  &   1968--2005 & 628 &  $8.58 \, 10^{14} $  &  $ 2.7 \, 10^{-2} $  &  
$ 3.7 \, 10^{-3} $  & 23.2 & 0.135 $\pm$ 0.004  &    $\ldots$    \\
          3000 \AA  & 1978--2005 &  210  &  $ 1.00 \, 10^{15} $  &  $ 2.5 \, 10^{-2} $  &  
$ 3.5 \, 10^{-3} $  & 25.0 & 0.140 $\pm$ 0.006  &    $\ldots$    \\
          2700 \AA  &  1978--1996 &  199  &  $ 1.11 \, 10^{15} $  &  $ 2.3 \, 10^{-2} $  &  
$ 3.5 \, 10^{-3} $  & 25.5 & 0.150 $\pm$ 0.008  &   $\ldots$   \\
          2425 \AA  &  1978--2005 &  210  &  $ 1.24 \, 10^{15} $  &  $ 2.0 \, 10^{-2} $  &  
$ 3.4 \, 10^{-3} $  & 24.5  & 0.167 $\pm$ 0.008  &    $\ldots$    \\
          2100 \AA  &  1978--2005 &  191  &  $ 1.43 \, 10^{15} $  &  $ 1.8 \, 10^{-2} $  &  
$ 3.4 \, 10^{-3} $  & 25.7 & 0.18 $\pm$ 0.01  &   $\ldots$   \\
          1950 \AA  &  1978--1996 &  237  &  $ 1.54 \, 10^{15} $  &  $ 1.9 \, 10^{-2} $  &  
$ 3.5 \, 10^{-3} $  & 29.3 & 0.177 $\pm$ 0.009  &   $0.5 {+1.2 \atop -0}$  \\
          1700 \AA  &  1978--1996 &  238  &  $ 1.76 \, 10^{15} $  &  $ 1.7 \, 10^{-2} $  &  
$ 3.3 \, 10^{-3} $  & 29.9 & 0.20 $\pm$ 0.01  &   $0.5 {+1.2 \atop -0}$  \\
          1525 \AA  &  1978--1996 &  239  &  $ 1.97 \, 10^{15} $  &  $ 1.6 \, 10^{-2} $  &  
$ 3.1 \, 10^{-3} $  & 31.5  & 0.193 $\pm$ 0.010  &   $0.5 {+0.4 \atop -0}$  \\
          1300 \AA  &  1978--1996 &  235  &  $ 2.31 \, 10^{15} $  &  $ 1.2 \, 10^{-2} $  &  
$ 2.6 \, 10^{-3} $  & 27.8  & 0.21 $\pm$ 0.01  &   $0.5 {+0.1 \atop -0.2}$  \\
         0.1 keV  & 1990--1995 &   20  &  $ 2.42 \, 10^{16} $  &  $ 2.7 \, 10^{-4} $  &  
$ 7.1 \, 10^{-5} $  & 6.5 & 0.24 $\pm$ 0.06  &   $\ldots$   \\
         0.2 keV  &  1990--1993 &  17  &  $ 4.84 \, 10^{16} $  &  $ 1.1 \, 10^{-4} $  &  
$ 1.7 \, 10^{-5} $  & 5.3 & 0.16 $\pm$ 0.03  &   $\ldots$   \\
         0.5 keV  &  1979--2005 &  58  &  $ 1.21 \, 10^{17} $  &  $ 3.1 \, 10^{-5} $  &  
$ 6.3 \, 10^{-6} $  & 3.8 & 0.20 $\pm$ 0.02  &   $\ldots$   \\
           1 keV  &  1978--2005 &  69  &  $ 2.42 \, 10^{17} $  &  $ 1.6 \, 10^{-5} $  &  
$ 3.6 \, 10^{-6} $  & 3.9 & 0.22 $\pm$ 0.02  &   $\ldots$   \\
           2 keV  &  1974--2005 &  93  &  $ 4.84 \, 10^{17} $  &  $ 8.7 \, 10^{-6} $  &  
$ 2.6 \, 10^{-6} $  & 4.2 & 0.27 $\pm$ 0.02  &   $\ldots$   \\
           5 keV  &  1970--2005 & 1032  &  $ 1.21 \, 10^{18} $  &  $ 4.7 \, 10^{-6} $  &  
$ 1.2 \, 10^{-6} $  & 5.7  & 0.249 $\pm$ 0.006  &    $\ldots$    \\
4--9 keV         & 1996--2005 & 2567  &  $ 1.57 \, 10^{18} $  &    6.2$^a$  &   1.4$^a$  & $\ldots$  &
0.230 $\pm$ 0.003  & $0.216 {+0.003 \atop -0.04}$  \\
          10 keV  & 1974--2005 & 1026  &  $ 2.42 \, 10^{18} $  &  $ 2.9 \, 10^{-6} $  &  
$ 8.2 \, 10^{-7} $  & 7.0  & 0.265 $\pm$ 0.007  &    $\ldots$    \\
9--20 keV        & 1996--2005 & 2567  &  $ 3.51 \, 10^{18} $  &    2.3$^a$  &   0.6$^a$  & $\ldots$  &
0.239 $\pm$ 0.004  &   $0.221 {+0.002 \atop -0.003}$  \\
          20 keV  & 1976--2005 &  987  &  $ 4.84 \, 10^{18} $  &  $ 1.8 \, 10^{-6} $  &  
$ 5.4 \, 10^{-7} $  & 8.7 &  0.275 $\pm$ 0.008  &    $\ldots$    \\
20--70 keV       & 1991--2000 &  91  &  $ 1.09 \, 10^{19} $  &    2.6$^b$  &   1.2$^b$  & $\ldots$ &
0.41 $\pm$ 0.04  &   $1.3 {+0.4 \atop -0.3}$  \\
          50 keV  & 1977--2005 & 1003  &  $ 1.21 \, 10^{19} $  &  $ 9.6 \, 10^{-7} $  &  
$ 3.6 \, 10^{-7} $  & 11.6  &  0.33 $\pm$ 0.02  &    $\ldots$    \\
         100 keV  & 1978--2005 & 1001  &  $ 2.42 \, 10^{19} $  &  $ 5.9 \, 10^{-7} $  &  
$ 2.3 \, 10^{-7} $  & 14.3 & 0.34 $\pm$ 0.01  &    $\ldots$    \\
         200 keV  & 1978--2005 &   41  &  $ 4.84 \, 10^{19} $  &  $ 4.8 \, 10^{-7} $  &  
$ 3.2 \, 10^{-7} $  & 23.2 & 0.40 $\pm$ 0.08  &   $\ldots$   \\
70--430 keV      & 1991--2000 &  91  &  $ 6.04 \, 10^{19} $  &    4.2$^b$  &   2.2$^b$  & $\ldots$  &
0.47 $\pm$ 0.05  &   $1.4 {+0.3 \atop -0.6}$  \\
         500 keV  & 1990--1999 &   15  &  $ 1.21 \, 10^{20} $  &  $2.8 \, 10^{-7} $  &  
$ 1.7 \, 10^{-7} $  & 33.9 & 0.43 $\pm$ 0.09  &   $\ldots$   \\
           1 MeV  & 1990--1999 &   20  &  $ 2.42 \, 10^{20} $  &  $ 1.7 \, 10^{-7} $  &  
$ 9.9 \, 10^{-8} $  & 41.1  &   $\ldots$   &   $\ldots$   \\
          30 MeV  & 1991--1999 & 17  &  $7.25 \, 10^{21} $  &  $3.3 \, 10^{-9} $  &  
$ 2.5 \, 10^{-9} $  & 23.9 & 0.58 $\pm$ 0.15  &   $\ldots$   \\
         100 MeV  & 1976--1997 &   28  &  $ 2.42 \, 10^{22} $  &  $ 3.6 \, 10^{-10} $  &  
$ 2.6 \, 10^{-10} $  & 8.7 & 0.64 $\pm$ 0.07  &   $\ldots$   \\
         300 MeV  & 1976--1997 & 16  &  $7.25 \, 10^{22} $  &  $8.3 \, 10^{-11} $  &  
$ 5.5 \, 10^{-11} $  & 6.0 & 0.48 $\pm$ 0.13  &   $\ldots$   \\
           1 GeV & 1991--1997 &  15  &  $2.42 \, 10^{23} $  &  $1.4 \, 10^{-11} $  &  
$ 9.4 \, 10^{-12} $  & 3.4 & 0.20 $\pm$  0.17  &   $\ldots$   \\
           3 GeV  & 1991--1997 &  15  &  $7.25 \, 10^{23} $  &  $2.9 \, 10^{-12} $  &  
$ 2.2 \, 10^{-12} $  & 2.1 &  $\ldots$  &   $\ldots$   \\
          10 GeV  & 1991--1997 & 14  &  $2.42 \, 10^{24} $  &  $5.7 \, 10^{-13} $  &  
$ 5.3 \, 10^{-13} $  & 1.4 &  $\ldots$  &  $\ldots$   \\

\end{longtable}

\end{document}